\documentclass[prb,twocolumn,superscriptaddress,showpacs,amsmath,floatfix]{revtex4}
\usepackage{graphicx}

\newcommand{\lb}{\left(}
\newcommand{\rb}{\right)}

\newcommand{\lcb}{\left[}
\newcommand{\rcb}{\right]}

\newcommand{\lcrb}{\left\{}
\newcommand{\rcrb}{\right\}}

\newcommand{\ld}{\left\vert}
\newcommand{\rd}{\right\vert}

\newcommand{\ggr}{\check{g}}

\newcommand{\gr}{\hat{g}}
\newcommand{\grR}{\hat{g}^R}
\newcommand{\grA}{\hat{g}^A}
\newcommand{\grK}{\hat{g}^K}
\newcommand{\grRA}{\hat{g}^{R(A)}}

\newcommand{\fR}{\hat{f}^{R}}

\newcommand{\fRA}{\hat{f}^{R(A)}}

\newcommand{\tr}{\operatorname{Tr}}
\newcommand{\sign}{\operatorname{sgn}}

\newcommand{\tid}{\hat{\tau}_{0}}
\newcommand{\tx}{\hat{\tau}_{1}}
\newcommand{\ty}{\hat{\tau}_{2}}
\newcommand{\tz}{\hat{\tau}_{3}}

\newcommand{\ctz}{\check{\tau}_{3}}

\newcommand{\ep}{\varepsilon}
\newcommand{\kep}{\kappa_{\varepsilon}}

\date{\today}

\begin{document}

\title{Nonlinear Resonance of Superconductor/Normal Metal Structures to Microwaves}

\author{E.~Kandelaki}
\affiliation{Theoretische Physik III,
Ruhr-Universit\"{a}t Bochum, D-44780 Bochum, Germany}

\author{A.~F.~Volkov}
\affiliation{Theoretische Physik III,
Ruhr-Universit\"{a}t Bochum, D-44780 Bochum, Germany}
\affiliation{Institute of Radioengineering and Electronics of the Russian Academy\\
of Sciences, 103907 Moscow, Russia}

\author{K.~B.~Efetov}
\affiliation{Theoretische Physik III,
Ruhr-Universit\"{a}t Bochum, D-44780 Bochum, Germany}

\author{V.~T.~Petrashov}
\affiliation{Department of Physics, Royal Holloway, University of London,\\
Egham, Surrey TW20 0EX, United Kingdom}

\begin{abstract}
We study the variation of the differential conductance $G=dj/dV$ of a normal
metal wire in a Superconductor/Normal metal heterostructure with a cross
geometry under external microwave radiation applied to the superconducting
parts. Our theoretical treatment is based on the quasiclassical Green's
functions technique in the diffusive limit. Two limiting cases are
considered: first, the limit of a weak proximity effect and low microwave
frequency, second, the limit of a short dimension (short normal wire) and
small irradiation amplitude.
\end{abstract}


\maketitle

\section{Introduction}

Superconductor/normal metal (S/N) nanostructures, where the
proximity effect (PE) plays an important role, have been studied very
actively during last two decades. Interesting phenomena have been discovered
in the course of these studies. Perhaps, the most remarkable one is an
oscillatory dependence of the conductance of a normal wire attached to two
superconductors which are incorporated into a superconducting
loop\cite{Petr92,Petr93}. This phenomenon was observed in the so-called
``Andreev interferometers'', i.e. in multi-terminal SNS junctions
(see \cite{Petr93,Pothier94,Vegvar94,Klapwijk95,Chandra98} as well as
reviews \cite{Been,LambRaim,NazRev,Belzig} and references therein).
The reason for this oscillatory behavior of the differential conductance $G=dj/dV$
is a modification of the transport properties of the $n$ wire due to the PE, i.e. due
to the condensate induced in the $n$ wire. The density of the induced
condensate is very sensitive to an applied magnetic field $H$ and oscillates
with increasing $H$.\\

Theory\cite{VZK,Hekking93,Zaitsev94,Nazarov96} was successful in
explaining the experiments and predicting new phenomena, including the
re-entrance of the conductance to the normal state in mesoscopic proximity
conductors\cite{AVZ79,Nazarov96,LambVolkov96} and transitions to the
$\pi$-state in the voltage-biased Andreev interferometers due to non-equilibrium
effects\cite{Volkov95,Schon98,Yip98}. The non-monotonic behavior of the
conductance in SN point contacts and controllable nanostructrures has been
observed in Refs.\cite{Marg79,Pannetier96,Petr96}, and the change of the
sign of the critical Josephson current in multiterminal SNS junctions has
been found in Refs.~\cite{KlapWees,Petr00}. Many important
results of the study of the SN mesoscopic structures are reviewed in
Refs.~\cite{Been,LambRaim,NazRev,Belzig}.\\

The so-called $\pi$-states have also been realized in equilibrium Josephson
SFS junctions with a ferromagnetic (F) layer between
superconductors\cite{GolubovRMP,BuzdinRMP,BVERMP} or in SIS Josephson junctions
of high-$T_{c}$, d-wave superconductors\cite{Kirtley,vHarlingen}.\\

A number of new phenomena have been discovered in thin one-dimensional N and
S wires\cite{Tinkham00,Arut05,Chan05} (see also~\cite{Arut08} for a recent
review and references therein).\\

Mesoscopic SNS structures proved to be a promising alternative to
Superconducting Quantum Interference Devices (SQUIDs) for certain
applications, including magnetic flux measurements and read-out of quantum
bits (qubits)\cite{Petr05} with a potential to achieve higher than
state-of-the-art fidelity, sensitivity and read-out speed. To achieve such
challenging aims extensive investigations of high frequency properties of
S/N nanostructures on a scale similar to that of SQUIDs are in order.\\

Studies undertaken to date concerned mainly the stationary properties of S/N
structures. Experimental data on S/N structures under microwave radiation
appeared only recently\cite{Aprili09,Petr10}. As to theoretical studies, one
can mention two papers\cite{VZK,Zaikin09} where the ac impedance of a S/N
structure was calculated. However, measuring the frequency dependence of the
ac conductance is not an easy task. It is more convenient to measure a
nonlinear dc response (dc conductance) to a microwave radiation. Recently, a
numerical calculation of the dependence of the critical Josephson current
$I_{c}$ in SNS junction on the amplitude of an external ac radiation has been
performed\cite{Bergeret10}.\\

In this paper, using a simple model we calculate the dc conductance of a
normal ($n$) wire in an S/N structure (cross geometry) as a function of the
frequency $\Omega $ and the amplitude of the external microwave radiation.
We consider the limiting cases of a long and a short $n$ wire and show that
the response has a resonance peak at a frequency $\Omega $ close to
$\ep_{s}/\hbar $, where $\ep_{s}$ is the energy of a subgap
in the $n$ wire induced by the PE. Our theory predicts novel resonances and
can help to optimize quantum devices based on hybrid SNS
nanostructures\cite{Petr05,Pekola}.\\

We employ the quasiclassical Green's function technique in the diffusive
limit. This means that we will solve the Usadel equation\cite{Usadel} for
the retarded (advanced) Green's function $\hat{g}^{R(A)}$ and the
corresponding equation for the Keldysh matrix function $\hat{g}^{K}$
(section 2). First, a weak PE will be considered when the Usadel equation
can be linearized (section 3). We calculate the dc conductance of the $n$
wire in this limit, assuming that the frequency of the ac radiation is low
($\Omega \ll T$). In section 4, the opposite limiting case of a short $n$ wire
will be analyzed for arbitrary frequencies $\Omega $. We present the
frequency dependence of the correction to the dc conductance caused by ac
radiation. In section 5, we discuss the obtained results.

\section{Model and Basic Equations}
\begin{figure}
  \includegraphics[height = 5.4 cm]{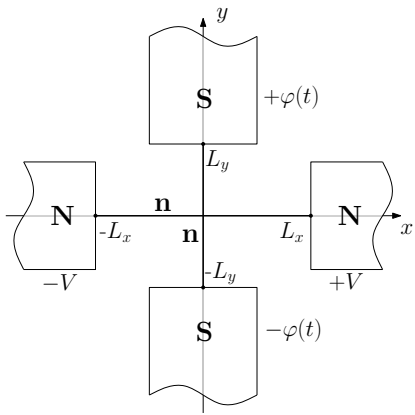}\\
  \caption{Structure under consideration}
  \label{fig:structure}
\end{figure}
We consider an S/N structure shown in Fig.~\ref{fig:structure}.
It consists of a $n$ wire or $n$\ film which connects two $N$ and $S$
reservoirs ($n$ and $N$ stand for a normal metal, $S$ means a superconductor).
The superconducting reservoirs may be connected by a superconducting contour.
The transverse dimensions of the $n$ wire are supposed to be smaller than
characteristic dimensions of the problem, but larger than the Fermi wave length
and the mean free path $l$ (diffusive case). This implies that all quantities depend
only on coordinates along the wire (the $x-$coordinate in the horizontal
direction and the $y-$coordinate in the vertical direction). The dc voltage
$2V$ is applied between the normal $N$ reservoirs, and the phase difference
$2\varphi $ exists between the superconducting reservoirs. The phase $\varphi$
is assumed to be time-dependent
\begin{align}
    \varphi(t) = \varphi_{0} + \varphi_{\Omega} \cos(\Omega t)  \label{Phase}
\end{align}
and related to the magnetic flux $\Phi$ inside the superconducting contour:
$\varphi(t) = \pi \Phi(t)/\Phi_{0}$ with $\Phi(t) = H(t)S$, where $H(t)$ is
an applied magnetic field and $S$ is the area of the superconducting
contour; that is, the magnetic field contains not only a constant component,
but also an oscillating one.\\

For simplicity, we assume the structure to be symmetric both in the
horizontal and vertical directions. This implies, in particular, that the
interface resistances $R_{nN}$ at $x=\pm L_{x}$ are equal to each other
(correspondingly, $R_{nS}(L_{y})=R_{nS}(-L_{y})$). Our aim is to calculate
the differential dc conductance $G$ between the $N$ reservoirs
\begin{equation}
    G = \frac{dj}{dV}  \label{G}
\end{equation}
as a function of the amplitude of the ac signal $\varphi_{\Omega}$ and the
frequency $\Omega$.\\

The calculations will be carried out on the basis of the well developed
quasiclassical Green's functions technique (see the reviews\cite
{LO,RS,Belzig,Kopnin}) which successfully was applied to the study
of $S/N$ structures (see for example%
\cite{VZK,LambRaim,NazRev,Belzig,VKogan,Volkov&Pavl,Zaikin09,Zaikin97,ArtVolkov}).
In this technique, all types of Green's functions (the ''normal''
and Gor'kov's functions as well as the retarded, advanced and Keldysh
functions) are matrix elements of a $4\times 4$ matrix
\begin{equation}
	\ggr =
		\begin{pmatrix}
			\grR & \grK \\
			0    & \grA
		\end{pmatrix}
			\label{gMatrix}
\end{equation}
where $\grRA$ are matrices of the retarded (advanced)
Green's functions, and $\grK$ is a matrix of the Keldysh
functions. The first matrices describe thermodynamical properties of the
system (the density of states, supercurrent etc), whereas the matrix
$\grK$ is used to describe dissipative transport and nonequilibrium
properties.\\

The matrix $\ggr$ satisfies the normalization condition\cite{LO75}
\begin{equation}\label{eq:norm_check_time}
    \lb \ggr {\,\circ\,} \ggr \rb (t,t') = \delta(t-t')
\end{equation}
where "${\circ}$" denotes the integral product
$\lb \ggr {\,\circ\,} \ggr \rb (t,t') = \int dt_1\, \ggr(t,t_1) {\cdot} \ggr(t_1,t')$
and "${\cdot}$" is the conventional matrix product. The Fourier transform performed
as $\ggr(\ep,\ep') = \int dt\, dt'\, e^{i\ep t - i \ep' t'} \ggr(t,t')$
yields $ \lb \ggr {\,\circ\,} \ggr \rb (\ep,\ep') = 2\pi \delta(\ep-\ep')$
where now $\lb \ggr {\,\circ\,} \ggr \rb (\ep,\ep') =
\int \frac{d\ep_1}{2\pi}\, \ggr(\ep,\ep_1) {\cdot} \ggr(\ep_1,\ep')$.\\

The matrix of Keldysh functions $\grK$ can be expressed in terms of the matrices $\grRA$
and a matrix of distribution functions $\hat{F}$:
\begin{equation}
	\grK = \grR {\circ} \hat{F} - \hat{F} {\circ} \grA
			\label{Keldysh}
\end{equation}
where the matrix $\hat{F}$ can be assumed to be diagonal~\cite{Belzig}:
\begin{align}
	\hat{F} & = \tid F_{+} + \tz F_{-}.
\end{align}
Here $\tid$ is the identity matrix and $\tz$ the third Pauli matrix. The function $F_{-}$ describes
the charge imbalance (premultiplied with the DOS and integrated over all energies
it gives the local voltage), while $F_{+}$ characterizes the energy distribution of quasiparticles.\\

Due to the general relation\cite{RS}
\begin{equation}
    \grA(\ep,\ep') = - \tz \grR{}^\dagger(\ep',\ep) \tz
\end{equation}
one can immediately calculate $\grA$ after finding the matrix $\grR$. That means that
knowing the matrices $\grR$ and $\hat{F}$ we can determine all entries of $\ggr$.\\

The Green's functions in $N$ and $S$ reservoirs are assumed to have an
equilibrium form corresponding to the voltages $\pm V$ and phases
$\pm \varphi(t)$. For example, the retarded (advanced) Green's functions
in the upper $S$ reservoir are
\begin{equation}
    \grRA_{S}(t,t') = \hat{S}(t) {\cdot} \grRA_{S0}(t-t') {\cdot}
        \hat{S}^{\dagger}(t')
            \label{g-inS}
\end{equation}
where $\hat{S}(t)=\exp [i\tz\varphi(t)/2]$ is a unitary transformation matrix
and the Fourier transform of $\grRA_{S0}(t-t')$ equals
\begin{equation}
	\grRA_{S0}(\ep) = \frac{1}{\xi_{\ep}^{R(A)}}
		\begin{pmatrix}
			\ep     & \Delta \\
			-\Delta & -\ep
		\end{pmatrix}
			\label{g0}
\end{equation}
with $\xi_{\ep}^{R(A)}=\pm\sqrt{(\ep \pm i0)^{2}-\Delta^{2}}$, i.e. the matrix
$\grRA_{S0}$ describes the BCS superconductor in the absence of phase.
The retarded (advanced) Green's functions in the lower $S$ reservoir are
determined in the same way with the replacement $\varphi(t) \rightarrow -\varphi(t)$.
The matrix $\grRA_{N}$ in the right (left) $N$ reservoirs is equal to
$\grRA_{S0}$ with $\Delta=0$, i.e. $\grRA_{N}=\pm\tz$.\\

In the reservoirs the matrix $\hat{F}(t,t')$
can be represented through the equilibrium distribution
$F_{eq}(\ep)=\tanh (\ep /2T)$ via Eq.~\eqref{g-inS}
\begin{equation}
		\hat{F}(t,t') = \hat{S}(t) {\cdot} F_{eq}(t-t') \hat{S}^{\dagger}(t').
				\label{F(t-t')}
\end{equation}
The phase $\varphi (t)$ in the upper $S$ reservoir is given by Eq.~\eqref{Phase},
and for $\varphi_N(t)$ in the right $N$ reservoir, we have: $\varphi_N
(t)=2eVt$. Therefore in the normal reservoir (at the right) the matrix
distribution function has diagonal elements
$\hat{F}_{N}(\ep)_{11,22}=\tanh \lb \frac{1}{2T} \lb \ep \pm eV\rb \rb$,
and can be written as $\hat{F}_{N}(\ep) = \tid F_{N+}(\ep ) + \tz F_{N-}(\ep)$
\begin{equation}
	F_{N\pm}(\ep) = \frac {1}{2}
				\lcb \tanh \frac{\ep + eV}{2T} \pm \tanh \frac{\ep - eV}{2T} \rcb. \label{F-inN}
\end{equation}
Thus, all Green's functions in the reservoirs are defined above.\\

Our task is to find the matrix $\ggr$ in the $n$ wire. In the considered diffusive
limit it obeys the equation\cite{LO75}
\begin{equation}
		\ctz {\cdot} \frac{\partial\ggr}{\partial {t}} +
            \frac{\partial\ggr}{\partial {t'}} {\cdot} \ctz+
				i(eV_{n}(t) \ggr - \ggr eV_{n}(t')) -
					D\nabla (\ggr  {\circ} \nabla \ggr) = 0,
							\label{EqSuper}
\end{equation}
where $\ctz$ is a diagonal matrix with equal elements
$(\ctz)_{11,22}=\tz$, $V_{n}$ is a local electrical potential in the $n$ wire.
We dropped the inelastic collision term supposing that \mbox{$E_{Th}=D/L_{max}^{2}\gg \tau_{inel}^{-1}$,}
where $D$ is the diffusion coefficient, \mbox{$L_{max}=\max \{L_{x,y}\}$} and $\tau_{inel}$ is
an inelastic scattering time. This equation is complemented by the
boundary \mbox{condition\cite{KL}}
\begin{equation}
	\ggr {\,\circ\,} \partial_{x,y}\ggr |_{x,y = \pm L_{x,y}} =
		\pm \kappa_{N,S} [\ggr ,\ggr _{N,S}]_{\circ}
			\label{BC}
\end{equation}
where $\kappa_{N,S}=1/(2 \sigma R_{nN,nS}), R_{nN,nS}$ are the $nN$ and $nS$
interface resistances per unit area and $\sigma $ is the conductivity of the
$n$ wire. Here we introduced the commutator $[\ggr ,\ggr_{N,S}]_{\circ} =
\ggr {\,\circ\,} \ggr_{N,S} - \ggr_{N,S} {\,\circ\,} \ggr$. The current in the $n$ wire
is determined by the formula
\begin{equation}
	j = \frac{\sigma}{8e} \tr \lcrb \tz {\cdot} 2\pi
            \lb \ggr {\,\circ\,} \partial_{x} \ggr \rb_{12} \lb t,t \rb \rcrb
					\label{Current}
\end{equation}
The matrix element $ \lb \ggr {\,\circ\,} \partial_{x}\ggr \rb_{12}$ is the
Keldysh component that equals $\lb \ggr {\,\circ\,} \partial_{x}\ggr \rb_{12} =
\grR {\,\circ\,} \partial_{x} \grK + \grK {\,\circ\,} \partial_{x} \grA$.\\

Even in a time-independent case, an analytical solution of the problem can
by found only under certain assumptions\cite{LambRaim,NazRev,Belzig,
Volkov&Pavl,VKogan,VZK,Zaikin09,Zaikin97}. In the considered case of a
time-dependent phase difference, the problem becomes even more complicated.
In order to solve the problem analytically, we consider two limiting cases:
a) weak proximity effect and slow phase variation in time; b) strong
proximity effect in a short $n$ wire and arbitrary frequency $\Omega$ of
the phase oscillations.

\section{Weak Proximity Effect; Slow Phase Variation}

In this section we will assume that the proximity effect is weak and the
phase difference $\varphi (t)$ is almost constant in time. The
latter assumption means that the frequency of phase variation satisfies the
condition $\Omega \ll T/\hbar $.
The weakness of the PE means that the anomalous (Gor'kov's) part $\fRA$
of the retarded and advanced Green's functions in the $n$ wire
$\grRA = g^{R(A)} \tz + \fRA$ can be assumed to be small
\begin{equation}
		\vert \fRA \vert \ll 1.  \label{smallness}
\end{equation}
The matrix $\fRA$ contains only off-diagonal elements. The diagonal part
obtained from the normalization is
\begin{equation}
		g^{R(A)} \tz \approx \pm\, \tz \lb 1 - \frac12 \lb \fRA \rb^2\rb.  \label{eq:gRA_by fRA}
\end{equation}
Now we can
linearize Eq.~\eqref{EqSuper} for the component $11(22)$, that is, for
the retarded (advanced) Green's functions. Then we obtain a simple linear
equation
\begin{equation}
    \nabla^{2} \fRA - \kep^2 \fRA = 0
        \label{WPEusadel}
\end{equation}
where $\kep^{R(A)}=\sqrt{\mp 2i\ep / D}$. The boundary conditions (\ref{BC}) for the matrices $\fRA$ acquire
the form
\begin{gather}
    \lcb \partial_{x} \fRA + 2 \kappa_{N} \hat{f}^{R(A)}\rcb\Big\vert_{x=+L_x} \!\!= 0;\\
    \lcb \partial_{x} \fRA - 2 \kappa_{N} \hat{f}^{R(A)}\rcb\Big\vert_{x=-L_x} \!\!= 0;\\
    \!\!\!\!\lcb \partial_{y} \fRA - 2 \kappa_{S} \lb \fRA_{S,+\varphi} \mp g^{R(A)}_{S} \!\! \cdot \! \fRA \rb\rcb\Big\vert_{y=+L_y}\!\!=0;\\
    \!\!\!\!\lcb \partial_{y} \fRA + 2 \kappa_{S} \lb \fRA_{S,-\varphi} \mp g^{R(A)}_{S} \!\! \cdot \! \fRA \rb\rcb\Big\vert_{y=-L_y}\!\!=0.
\end{gather}
As follows from Eq.~\eqref{g-inS} the functions $g_{S}$, $\hat{f}_{S,\varphi}$ are
\begin{gather}
    g^{R(A)}_{S}=\ep/\xi_{\ep}^{R(A)},\label{eq:grRA_S}\\
    \fRA_{S,\varphi}= (i\ty\cos\varphi + i\tx\sin\varphi) \Delta/\xi_{\ep}^{R(A)}.\label{eq:fRA_S}
\end{gather}
We took into account that $\varphi(t)$ is almost constant in time.\\

One can show that the solution for $\fR$ in the horizontal part of the $n$ wire is
\begin{gather}
	\fR = i \ty f(x),\\
     f(x) = C\, \cosh(\theta_{x}x/L_{x}) + \sign(x)\, S\, \sinh(\theta_{x}x/L_{x})  \label{WPEfR}
\end{gather}
where $\sign(x)$ is the sign function. Dropping the index $R$ of the quantities
$\kep,g_S(\ep),\xi_{\ep}$ the integration constants $C$ and $S$ can be written as
\begin{subequations}\label{eq:WPE_CS}
    \begin{gather}
        C (\ep,\varphi) =
            \frac{ \lb \theta_x \cosh\theta_x + r_N \sinh\theta_x \rb \cdot r_{S} \Delta \cos\varphi / \xi_{\ep}}
            {\mathcal{D}(\ep)},\\
        S (\ep,\varphi) =
            -\frac{ \lb r_{N} \cosh \theta_{x} + \theta_{x} \sinh \theta_{x} \rb \cdot r_{S} \Delta\cos\varphi /\xi_{\ep}}
            {\mathcal{D}(\ep)}
    \end{gather}
\end{subequations}
where $\mathcal{D}(\ep)=
(r_S g_S(\ep) \theta_x + r_N \theta_y) \cosh(\theta_x + \theta_y) +
(r_S g_S(\ep) r_N + \theta_x \theta_y) \sinh(\theta_x + \theta_y)$,
$r_{N,S} = 2 \kappa_{N,S} L_{x,y}$, $\theta_{x,y} = \kep L_{x,y}$.\\

Knowing the function $\hat{f}^{R}(x)$, one can find the
correction to the conductance of the $n$ wire due to the PE. In order to
obtain the current, we take the $(12)$ component (the Keldysh component) of
Eq.~\eqref{EqSuper}, multiply this component by $\tz$ and take the
trace. In the Fourier representation we get (compare with Eq.~(2) of
Ref.~\cite{VKogan})
\begin{equation}
	M_{-}(\ep,\varphi,x)\partial_{x}F_{-}(x)=c(\ep ,\varphi ).
	\label{WPEeqF}
\end{equation}
where the function $M_{-}(\ep ,\varphi,x) = 1 + \tfrac14 \lb f(x) + f^\ast(x) \rb^2$
determines the correction to the conductivity caused by the PE and
$c(\ep ,\varphi )$ is an integration constant that is related to the
current:
\begin{align}\label{eq:currentj_by_c}
    j=\frac{\sigma}{2e}\int_{-\infty}^{\infty} d\ep\, c(\ep).
\end{align}
It is determined
from the boundary condition that can be obtained from Eq.~\eqref{BC}
\begin{equation}
	M_{-}(\ep,\varphi,x)\partial_{x}F_{-}(x) = c(\ep,\varphi)=\nu
	   [ F_{N-} - F_{-}(L_{x}) ].  \label{WPEbc}
\end{equation}
where $\nu (\ep,\varphi) = \Re \{ 1 + \tfrac12 f(L_{x})^{2}\}$ is the
density of states in the $n$ wire near the $nN$ interface. Finding
$F_{-}(L_{x})$ and $c(\ep)$ from Eq.~\eqref{WPEbc}, we obtain for
the current density (compare with Eq.~(13) of Ref.~\cite{VZK})
\begin{equation}
	j(\varphi) = \frac{1}{2e} \int_{-\infty}^{\infty } d\ep
        \frac{ F_{N-} } { R_{nN}/\nu + R_{n} \langle M(\ep,\varphi)^{-1} \rangle }  \label{WPEj}
\end{equation}
Here $F_{N-}$ is defined according to Eq.~\eqref{F-inN}, $R_{n}=L_{x}/\sigma$ is the
resistance of the $n$ wire of the length $L_{x}$ in the normal state,
and $\langle M(\ep,\varphi)^{-1} \rangle = (1/L_{x}) \int_{0}^{L_{x}} dx\, M_{-}(\ep,\varphi,x)^{-1}$.
The first term in the denominator is the $nN$ interface resistance and the second term is the
resistance of the $(0,L_{x})$ section of the $n$ wire modified by the PE.
The expressions for the DOS $\nu (\ep,\varphi )$ and the function
$\langle M(\ep,\varphi)^{-1}\rangle$ are given in appendix.\\

\begin{figure}
  \includegraphics[height = 5.6 cm]{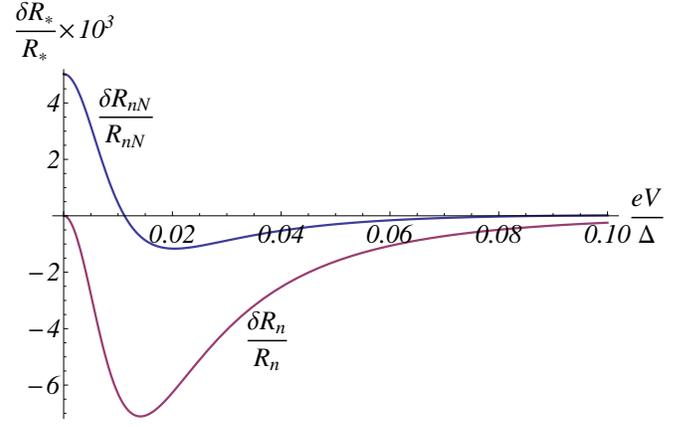}\\
  \caption{Bias voltage dependence of the normalized variations of the resistance
    contributions $\delta R_{nN}/R_{nN}$ and $\delta R_{n}/R_{n}$. Parameter values:
    $\varphi=\pi/3$, $L_y/L_x=1$, $\ep_{N}/\Delta=2.5 \cdot 10^{-2}$, $\ep_{S}/\Delta=5 \cdot 10^{-3}$, $R_{n}/R_{nN}=1$.
    }
  \label{fig:dRvsV}
\end{figure}

For the differential conductance $G = dj/dV$ at zero temperature we obtain
\begin{equation}
	{G}(V,\varphi (t))=
        \{ R_{nN}/\nu + R_{n}\langle M(eV,\varphi)^{-1}\rangle \}^{-1}.  \label{WPEGfin}
\end{equation}
In Fig.~\ref{fig:dRvsV} we show the dependence of the $nN$ interface resistance
variation $\delta R_{nN}=R_{nN}/\nu-R_{nN}$ and the resistance variation of the $n$ wire
$\delta R_{n}=R_{n}\langle M(eV,\varphi)^{-1}\rangle - R_{n}$ on the bias voltage $V$
for a fixed phase difference.
It can be seen that the $\delta R_{nN}$ is either positive or negative depending on $V$,
while $\delta R_{n}$ is always negative, i.e. the PE leads to voltage dependent changes of the
interface resistance (caused by the changes of the DOS in the $n$ wire) and to a
decrease of the resistance of the $n$ wire.\\

\begin{figure}
  \includegraphics[height = 5.6 cm]{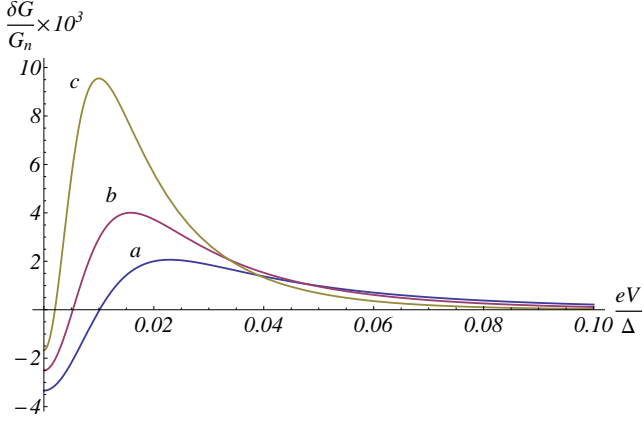}\\
  \caption{Bias voltage dependence of the normalized conductance variation $\delta G/G_{n}$.
    Parameter values: $\varphi=\pi/3$, $L_y/L_x=1$, $\ep_{N}/\Delta = 2.5 \cdot 10^{-2}$, $\ep_{S}/\Delta=5 \cdot 10^{-3}$.
    Different cases: a)~$R_{n}/R_{nN}=0.5$, b)~$R_{n}/R_{nN}=1$, c)~$R_{n}/R_{nN}=2$.
    }
  \label{fig:dGvsV}
\end{figure}

The conductance variation
$\delta G = {G}(V,\varphi)-{G}_{n}$, is shown in Fig.~\ref{fig:dGvsV} for various values
of $R_{nN}/R_{N}$, where ${G}_{n}=\{R_{nN}+R_{N}\}^{-1}$ is the conductance
of the $n$-wire in the normal state. These results have been obtained
earlier\cite{VZK,Nazarov96,Volkov&Pavl,Zaikin97}.\\

\begin{figure}
  \includegraphics[height = 5.6 cm]{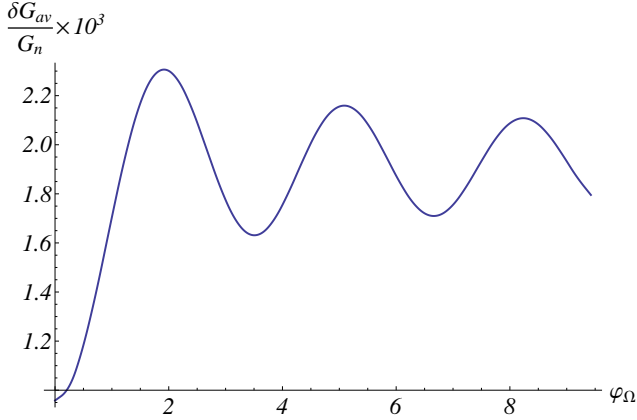}\\
  \caption{Bias voltage dependence of the normalized time-averaged conductance variation $\delta G_{av}/G_{n}$.
    Parameter values: $\varphi_0 = \pi/3$, $L_y/L_x = 1$, $\ep_{N}/\Delta = 2.5 \cdot 10^{-2}$, $\ep_{S}/\Delta = 5 \cdot 10^{-3}$,
    $eV/\Delta = 5 \cdot 10^{-2}$, $R_{n}/R_{nN}=1$.
    }
  \label{fig:dGav_vsphi}
\end{figure}

We are interested in the dc conductance variation averaged in time:
$\delta G_{av} = (\Omega/2\pi) \int_{0}^{2\pi/\Omega} dt\,\delta {G}(V,\varphi (t))$.
First, from Eqs.~(\ref{WPEfR}-\ref{eq:WPE_CS}) we can extract the dependence of
the function $f$ on the phase $\varphi$: $f(x,\varphi) = f(x,0)\cos\varphi$.
Hence we obtain
$M_{-}(\ep,\varphi,x) = 1 + \delta M_{-}(\ep,0,x) \cos^2\varphi$
where $\delta M_{-}(\ep,\varphi,x) = M_{-}(\ep,\varphi,x) - 1$.
At the same time, $\nu(\ep,\varphi) = 1 + \delta\nu(\ep,0)\cos^2\varphi$ with
$\delta \nu(\ep,\varphi) = \nu(\ep,\varphi) - 1$. These observations lead to
the relation
\begin{align}
    \delta G (V,\varphi(t)) = \delta G (V,0)\cos^2\varphi(t)
\end{align}
which by averaging over time yields
\begin{align}
    \delta G_{av} = \delta G (V,0) \cdot \frac12 \lb 1 + J_0 (2\varphi_\Omega) \cos(2\varphi_0) \rb
\end{align}
where $J_0$ is the Bessel function of the first kind and zeroth order. This oscillatory
behavior of the time-averaged (dc) conductance variation $\delta G_{av}$ as a function of the
ac amplitude can be seen in Fig.~\ref{fig:dGav_vsphi}.\\

Thus, the calculations carried out in this section under assumption
of adiabatic phase variations allow us to obtain the dependence of the
conductance change $\delta G_{av}$ on the amplitude $\varphi_{\Omega}$,
but provide no information about the frequency dependence of $\delta G_{av}$.
This dependence will be found in the next section.

\section{Strong PE; Short Normal Wire}

In this section we analyze the limiting case of a short $n$ wire when the
Thouless energy $E_{Th}=D/L_{x}^{2}$ is much larger than characteristic
energies: $E_{Th} \gg D\kappa _{N,S}^{2}, T, eV$. In this case all the
functions in Eq.~\eqref{EqSuper} are almost constant in space and we can
integrate this equation from $\{x,y\}=\pm 0$ to $\{x,y\} = \pm L_{x,y}$
over $x$ and $y$ coordinates. The term $\tz {\cdot} \partial_{t} \ggr +
\partial_{t'} \ggr {\cdot} \tz$ (in the Fourier representation
$- i \ep \tz {\cdot} \ggr + i \ep' \ggr {\cdot} \tz$) is considered as a constant
and the term with the voltage $V$ is omitted because we neglect the voltage
drop over the $n$ wire; the voltage drops across the $nN,nS$ interfaces and
is set to be zero in the $n$ wire. Performing this integration and summing
up the results, we obtain
\begin{gather}
	2(L_{x}+L_{y})\check{A} = \check{J}_{x}(L_{x})-\check{J}_{x}(-L_{x})-
	   \check{J}_{x}(+0)+\check{J}_{x}(-0)\nonumber\\
    +\check{J}_{y}(L_{y})-\check{J}_{y}(-L_{y})
        -\check{J}_{y}(+0)+\check{J}_{y}(-0)
    \label{SPE1}
\end{gather}
where $\check{J}_x (x)= D\, \ggr {\,\circ\,} \partial_{x}\ggr |_{y=0}$,
$\check{J}_y (y)= D\, \ggr {\,\circ\,} \partial_{y}\ggr |_{x=0}$,
and $\check{A} = -i(\ep \ctz{\cdot} \ggr - \ggr {\cdot} \ctz\ep ^{\prime })$.
Integration around the point $ ( x,y ) = ( 0,0 )$ yields the conservation of the ''currents''
(using terminology of the circuit theory\cite{NazRev})
\begin{equation}
	\check{J}_{x}(+0)+\check{J}_{y}(+0)=\check{J}_{x}(-0)+ \check{J}_{y}(-0) \label{SPE2}
\end{equation}
Combining Eqs.~(\ref{SPE1}-\ref{SPE2}) and the boundary conditions~\eqref{BC},
we arrive at the equation
\begin{equation}
	\ep \ctz{\cdot} \ggr -
        \ggr {\cdot} \ctz \ep' = i\ep_{N}
            [\ggr ,\ggr _{N+}]_{\circ} + i\ep_{S}[\ggr ,\ggr _{S+}]_{\circ}
                \label{SPE3}
\end{equation}
Here $\ep_{N,S}=D/(2 R_{nN,nS}\sigma L)$ is a characteristic energy
related to the interface transparencies, $L=L_{x}+L_{y}$. The energy
$\ep_{N}$ determines the damping in the spectrum of the $n$ wire and
the energy $\ep _{S}$ is related to a subgap induced in the $n$ wire
due to the PE. The matrices $\ggr_{N,S+}$ are equal to:
$\ggr_{N,S+} = \tfrac12 [\ggr_{N,S}(L_{x,y})+ \ggr_{N,S}(-L_{x,y})]$.\\

In the limit of the short $n$ wire considered in this section, we need to
find only the retarded (advanced) Green's functions. Indeed, let us rewrite
the expression for the current~\eqref{Current} using the boundary
condition~\eqref{BC} at the right $nN$ interface and concentrating on the
dc component of the current:
\begin{equation}
	j=\frac{1}{16 e R_{nN}}
        \tr\{ \tz{\cdot}\!\!\! \int\limits_{-\infty}^{\infty} \!\!\! d\ep \,
            (\grR \!\! {\cdot} \grK_{N} + \grK \!\! {\cdot} \grA_{N} - \grR_{N} {\cdot} \grK - \grK_{N} {\cdot} \grA)\}  \label{SPEcurrent}
\end{equation}
where $\grRA_{N}=\pm \tz$ and $\tr \{\tz {\cdot} (\grR {\cdot} \grK_{N})\} = 4 g^{R} F_{N-}$.
The distribution function $F_{N-}$ in the $N$ reservoir is defined in
Eq.~\eqref{F-inN}. The integral over energies from the second and third term is
zero because it is proportional to the voltage in the $n$ wire which is set
to be zero. Therefore the current can be written as
\begin{equation}
	j=\frac{1}{2eR_{nN}}\int\limits_{-\infty}^{\infty} d\ep\, \nu (\ep )F_{N-}(\ep)
	\label{SPEcurrent1}
\end{equation}
where $\nu(\ep) = \tfrac12 (g^{R}-g^{A}) = \Re\{g^{R}(\ep)\}$. This
formula has an obvious physical meaning - the current through the $nN$
interface is determined by the product of the DOS in the $n$ wire and $N$
reservoir ($\nu_{N}=1$) and the distribution function in the $N$ reservoir
(the distribution function $F_{-}$ in the $n$ wire is zero).\\

Using Eqs.~\eqref{G},~\eqref{F-inN},~\eqref{SPEcurrent1} we arrive at the following expression for the differential
conductance:
\begin{gather}\label{eq:G_SNW}
    G = \frac{1}{2 R_{nN}}\!\!\int\limits_{-\infty}^{\infty} \!\! \frac{d\ep}{4\,T} \nu (\ep)
        \! \lcb \frac{1}{\cosh^{2} \tfrac{\ep + eV}{2T}} + \frac{1}{\cosh^{2} \tfrac{\ep - eV}{2T}} \rcb
\end{gather}

In order to find the matrix $\grR$, we can write the (11) component of Eq.~\eqref{SPE3}
in the form
\begin{equation}
	\tilde{\ep} \tz {\cdot} \grR - \grR {\cdot} \tz \tilde{\ep}' =
        i\ep _{S}[\grR,\grR_{S+}]_{\circ}  \label{SPE4}
\end{equation}
where $\tilde{\ep}=\ep + i\ep _{N}$, $\tilde{\ep}' = \ep' + i\ep_{N}$.\\

According to Eqs.~\eqref{Phase},~\eqref{g-inS} the matrix $\grR_{S+}$ is a function of
two times, $\grR_{S+}(t,t')$, that is, in the Fourier representation it is function
of two energies: $\ep, \ep'$. Therefore, to find the matrix $\grR (\ep,\ep')$ in a
general case is a formidable task.\\

However, we can assume that the amplitude of the ac component of the phase $\varphi_{\Omega}$ is
small and obtain the solution making an expansion in powers of $\varphi_{\Omega}$:
\begin{equation} \label{eq:grR_expansion}
    \grR = \grR_0 + \delta_1 \grR + \delta_2 \grR + \ldots
\end{equation}
Here and later all matrix Green's functions written without arguments are functions of two energies
$(\ep,\ep')$. Those of them which are diagonal in energy may be also (explicitly) written with
a single energy argument, e.g. $\grR_{S0+}=\grR_{S0+}(\ep) 2\pi \delta(\ep-\ep')$.\\

Similar to Eq.~\eqref{eq:grR_expansion} we represent the matrix $\grR_{S+}$ (up to the second order in $\varphi_{\Omega}$)
as $\grR_{S+} = \grR_{S0+} + \delta_{1}\grR_{S+} + \delta_{2}\grR_{S+}$ and find from
Eq.~\eqref{g-inS} for the stationary part $\grR_{S0+}$ and the
corrections $\delta_{1}\grR_{S+}$ (first order in $\varphi_{\Omega}$) and $\delta_{2}\grR_{S+}$
(second order in $\varphi_{\Omega}$):
\begin{equation}
		\grR_{S0+} =  2\pi \delta_0 \lb \ep \tz + i \Delta \cos\varphi_0 \ty \rb \xi_{\ep}^{-1} \label{eq:grR_S0_PL}
\end{equation}
\begin{equation}
		\delta_{1} \grR_{S+} = - i \ty
            \tfrac{\pi}{2} \varphi_{\Omega} \Delta \sin \varphi_{0}
                (\xi_{\ep}^{-1} + \xi_{\ep'}^{-1} )
                    \lb \delta_{\Omega} + \delta_{-\Omega} \rb  \label{SPE5}
\end{equation}
\begin{equation}
    \delta_{2}\grR_{S+} =
        - i \ty \tfrac{\pi}{8} \varphi_{\Omega}^2 \Delta \cos\varphi_0 \lb P_0 + P_2 \rb \label{SPE6}
\end{equation}
where we used the notation $\delta_{\omega} \equiv \delta (\ep -\ep' + \omega)$, $\xi_{\ep}\equiv\xi_{\ep}^{R}$ and defined the functions
\begin{gather}
    P_0 = \delta_0 (2 \xi_{\ep'}^{-1} + \xi_{\ep'+\Omega}^{-1} + \xi_{\ep'-\Omega}^{-1}),\nonumber\\
    P_2 =  \tfrac12 \lb \delta_{2\Omega}  + \delta_{-2\Omega} \rb ( \xi_{\ep'}^{-1} + \xi_{\ep}^{-1} + 2 \xi_{\frac12(\ep+\ep')}^{-1}).
\end{gather}

Using the expressions for $\delta_{1}\grR_{S+}$ and $\delta_{2}\grR_{S+}$ given above we can calculate
the corrections to $\grR_{0}$ up to the second order in $\varphi_\Omega$ and the corresponding modification
of the DOS $\delta\nu$ in the $n$ wire.\\

In the zeroth-order approximation, i.e. for $\varphi_{\Omega}=0$ we obtain from
Eq.~\eqref{SPE4} $\grR_{0}(\ep,\ep')= \grR_{0}(\ep) 2\pi \delta(\ep-\ep')$ where
the matrix $\grR_{0}(\ep)$ obeys the equation
\begin{equation}\label{SPE7}
    [ \tz E_{\ep}^{R} + i \ty E_{sg}^{R}, \grR_{0}] = 0
\end{equation}
containing $E_{\ep}^{R} = \tilde{\ep} + i \ep_{S} g^{R}_{S0}(\ep) = \ep + i\ep_{N} + i \ep_{S} g^{R}_{S0}(\ep)$,
$E_{sg}^{R}=i \ep_{S} \cos \varphi_{0} f_{S0}^{R} (\ep)$,
$g^{R}_{S0}(\ep) = \ep/\xi^{R}_{\ep}$, $f^{R}_{S0}(\ep) = \Delta/\xi^{R}_{\ep}$.
The solution of this equation is~\cite{VZK}
\begin{gather}
		\grR_{0}(\ep) = \tz g^{R}_{0}(\ep) + i\ty f^{R}_{0}(\ep);\nonumber\\
        g^{R}_{0}(\ep) = E^{R}_{\ep}/\zeta^{R}_{\ep},\quad f^{R}_{0}(\ep) = E^{R}_{sg,\varphi}/\zeta^{R}_{\ep}  \label{SPE8}
\end{gather}
where $\zeta_{\ep}^{R}=\sqrt{\lb {E_{\ep }^{R}} \rb^{2} - \lb E_{sg}^{R} \rb^2}$.
The quantity $E_{sg}^{R}$ determines a subgap induced in the $n$ wire due
to the PE. Indeed, consider the most interesting case of small energies assuming that
$\{\ep,\ep_{S}\} \ll \Delta $; then, $\xi^{R}_{\ep} \approx i \Delta$, $f_{S0}^{R}(\ep)\approx -i$
and $\zeta_{\ep}^{R} \approx \sqrt{(\ep + i\ep_{N})^{2}-(\ep_{S}\cos\varphi_{0})^{2}}$.
This means that the spectrum of the $n$ wire has the same form as in the BCS superconductor
with a damping $\ep_{N}$ and a subgap $\ep_{S}\ld \cos\varphi_{0}\rd$,
which depends on the $nS$ interface transparency and phase difference.\\

Note that the formula for the subgap induced in the N metal due to
the PE in a tunnel SIN junction was obtained by McMillan\cite{McMillan}.\\

The obtained results for the functions $g_{0}^{R}(\ep)$ and
$f_{0}^{R}(\ep)$ can be easily generalized for the case of
asymmetric $nS$ interfaces with different interface resistances
$R_{nS1,2}$ (correspondingly, $\ep_{S1,2}$). In the limit
$\ep_{S1,2} \ll \Delta $, we obtain for the subgap $\ep_{sg}$
\begin{equation}
    \ep_{sg}(\varphi_0)=\tfrac12 \sqrt{ \ep_{S1}^{2}+\ep_{S2}^{2} + 2\ep_{S1}\ep_{S2}\cos 2\varphi_0 }  \label{SPE8a}
\end{equation}
This formula shows that that the subgap as a function of the phase
difference $\varphi$ varies from $\tfrac12|\ep_{S2}-\ep _{S1}|$
for $\varphi_0 = \pi/2$ to $\tfrac12(\ep _{S2}+\ep_{S1})$
for $\varphi_0 = 0$.\\

We proceed finding the corrections of the first ($\delta_{1}\grR$) and second
($\delta_{2}\grR$) order in $\varphi_{\Omega }$ for $\grR_0$ in a way similar to the one used
in\cite{ArtVolkov,Zaitsev99}. The correction of the first order $\delta_{1}\gr$
(for brevity, we drop the index $R$) obeys the equation
\begin{equation}
		\zeta_{\ep} \gr_{0}(\ep) {\cdot} \delta_{1} \gr - \delta_{1} \gr {\cdot} \gr_{0}(\ep') \zeta_{\ep'} =
            i \ep_{S} [\gr_{0}, \delta_{1}\gr_{S+}]_\circ \label{SPE9}
\end{equation}
which contains all terms of the first order in $\varphi_{\Omega}$ from Eq.~\eqref{SPE4}.
Note, that we are making use of the relation $\gr_0 (\ep) = \zeta_\ep^{-1} (\tilde{\ep}\tz + i \ep_S \gr_{S0+}(\ep))$
evident from Eqs.~\eqref{SPE4},~(\ref{SPE7}-\ref{SPE8}).\\

In order to solve Eq.~\eqref{SPE9}, it is useful to employ the normalization
condition~\eqref{eq:norm_check_time} for $\gr\equiv\grR$ which for the first-order term
of $\gr\circ\gr$ yields
\begin{gather}
    \gr_{0}(\ep) {\cdot} \delta_{1} \gr + \delta_{1}\gr {\cdot} \gr_{0}(\ep')=0 \label{SPE10}
\end{gather}
From Eqs.~(\ref{SPE9}-\ref{SPE10}), we find
\begin{equation}
		\delta_{1}\gr = i\ep_{S}
            \frac{\delta_{1}\ggr_{S+} - \gr_{0}(\ep ){\cdot} \delta_{1}\gr_{S+} {\cdot} \gr_{0}(\ep')}
                {\zeta_{\ep}+\zeta_{\ep'}}  \label{SPE11}
\end{equation}

We determine the correction $\delta_{2}\gr$ in the same manner.
Reading off the second-order terms in Eq.~\eqref{SPE4} gives
\begin{equation}
   [\zeta_{\ep} \gr_{0}, \delta_{2}\gr]_\circ = i\ep_{S}\Big( [\gr_{0}, \delta_{2}\gr_{S+}]_\circ + [\delta_{1}\gr, \delta_{1}\gr_{S+}]_\circ \Big)
\end{equation}
The second-order part of the normalization condition is
\begin{equation}
	\gr_{0}(\ep) {\cdot} \delta_{2}\gr+\delta_{2} \gr {\cdot} \gr_{0}(\ep')=
        -\delta_{1} \gr \circ \delta_{1}\gr  \label{SPE12}
\end{equation}
Thus, we obtain the second-order correction
\begin{align}
		\delta_{2}\gr & = i\ep _{S}\frac{\delta _{2}\ggr_{S+}-\gr_{0}(\ep ) {\cdot} \delta_{2}\gr_{S+}{\cdot} \gr_{0}(\ep')}
            {\zeta_{\ep} + \zeta_{\ep'}}\nonumber\\
        &\ + \left\{ i\ep_{S} \frac{[\delta_{1}\gr_{S+},\delta_{1}\gr]_{\circ}} {\zeta_{\ep}+\zeta_{\ep'}}
            - \frac{\zeta_{\ep} \lb \delta_{1}\gr \circ \delta_{1}\gr \rb} {\zeta_{\ep}+\zeta_{\ep'}} \right\}{\cdot} \gr_{0}(\ep').
		      \label{SPE13}
\end{align}

In order to calculate the correction to the dc conductance caused by
the ac radiation, $\delta G$, we need to find $\tr \{\tz {\cdot} \delta_{1} \hat{g}\}$
and $\tr \{\tz {\cdot} \delta_{2} \hat{g}\}$ and take their parts proportional to
$2\pi \delta(\ep-\ep')$. By inspection of Eqs.~\eqref{SPE5},~\eqref{SPE11} one
recognizes that the first-order correction contains only terms proportional to
$\delta(\ep - \ep' \pm \Omega)$ and therefore only contributes to the ac current.
This is the fundamental reason why the second-order analysis is needed to determine
the variation of the dc conductance.\\

As a result we just have to find the multiple of $2\pi \delta (\ep-\ep')$ contained in
$\tr \{\tz{\cdot} \delta_{2}\hat{g}\}$ which we denote as $2 \delta_{dc}g(\ep)$, that is
$\delta_{dc}g(\ep)2\pi \delta (\ep-\ep'):=\tfrac12 \tr \{\tz{\cdot} \delta_{2}\hat{g}\}_{dc}$.\\

We represent the function $\delta_{dc}g(\ep)$ as a sum
\begin{equation} \label{SPE13a}
    \delta_{dc} g(\ep) = \delta^{(0)}_{dc} g(\ep) + \delta^{(\Omega)}_{dc} g(\ep)
\end{equation}
where the function $\delta^{(0)}_{dc} g(\ep)$ originates from the first term
in Eq.~\eqref{SPE13} and the function $\delta^{(\Omega)}_{dc} g(\ep)$ from the
second and third terms. If we consider the case when the subgap
$\ep_{S} \ld \cos \varphi_{0} \rd$ is much less than $\Delta $, i.e.
\begin{equation}
    \ep_{S} \ld \cos \varphi_{0} \rd \ll \Delta  \label{SPE13b}
\end{equation}
then, at low energies $\ep \lesssim \ep_{S}$, the function
$\delta^{(0)}_{dc} g(\ep )$ is almost independent of $\Omega$, whereas
the function $\delta^{(\Omega)}_{dc} g(\ep)$ depends strongly on $\Omega$
at $\ep \approx \ep_{S} \ld \cos \varphi_{0} \rd$. Assuming the validity of
Eq.~\eqref{SPE13b} we obtain
\begin{equation}
    \delta^{(0)}_{dc}g(\ep ) = - \frac14 \ep_{S}^{2} \varphi_{\Omega}^{2} \cos^{2}\varphi_{0} \frac{g_{0}(\ep)}{\zeta^2_{\ep}}, \label{SPE14a}
\end{equation}
\begin{widetext}
    \begin{gather}
    		\!\!\!\!\!\!\delta^{(\Omega)}_{dc} g(\ep ,\Omega ) = \frac14 \ep_{S}^{2}\varphi_{\Omega}^{2} \sin^{2}\!\!\varphi_{0}
    			\sum_{\pm \Omega}
                    \frac{g_{0}(\ep) [1+f_{0}(\ep )f_{0}(\ep \!+\!\Omega )+g_{0}(\ep)g_{0}(\ep \!+\!\Omega )]}
    				    {[\zeta_{\ep} + \zeta_{\ep+\Omega}]^{2}} +
    				\frac{f_{0}(\ep )[g_{0}(\ep)f_{0}(\ep \!+\!\Omega )+f_{0}(\ep )g_{0}(\ep \!+\!\Omega )]}
    					{\zeta_{\ep}[\zeta_{\ep} + \zeta_{\ep +\Omega}]}
    									\label{SPE14b}
    \end{gather}
\end{widetext}
where the functions $g_{0}(\ep)$, $f_{0}(\ep)$, $\zeta_{\ep}$ are defined in
Eq.~\eqref{SPE8}. The sum sign index "$\pm \Omega$" in Eq.~\eqref{SPE14b}
means that the given expression is added to the same one with the negative frequency $(-\Omega)$.\\

Using the function $\delta_{dc}g(\ep,\Omega)$ we can calculate a
correction to the DOS $\delta\nu(\ep,\Omega)$ due to the PE and with
the aid of Eq.~\eqref{eq:G_SNW} find the correction $\delta G(V,\Omega)$
to the differential dc conductance. As follows from Eq.~\eqref{eq:G_SNW},
at zero temperature the normalized differential dc conductance
$\tilde {G}(V,\Omega) = G(V,\Omega) R_{nN}$ is equal to
\begin{equation}
    \tilde{G}(V,\Omega) \equiv \tilde{G}_{0}(V) + \delta \tilde{G}(V,\Omega) = \nu_{0}(eV) + \delta \nu (eV,\Omega)  \label{SPE15}
\end{equation}
with the definitions $\nu_{0}(eV) = \Re (g_{0}(eV))$ and $\delta \nu (eV,\Omega) = \Re(\delta_{dc} g(eV,\Omega))$.\\

\begin{figure}[htbp]
  \includegraphics[width = 8cm]{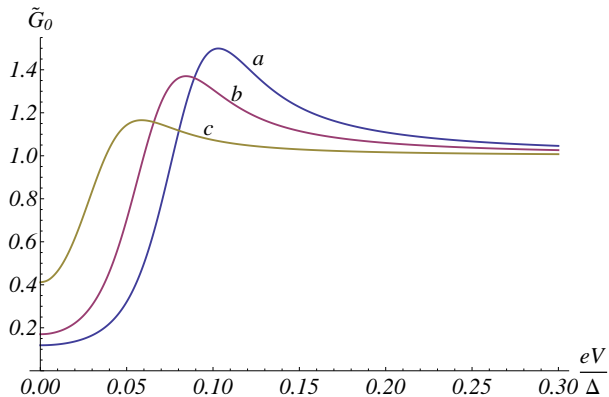}\\
  \caption{Normalized stationary differential conductance $\tilde{G}_0$ versus bias voltage $V$.
    Parameter values:
    $T/\Delta = 10^{-2}$, $\ep_S/\Delta=0.1$, $\ep_N/\Delta = 10^{-2}$.
    Different cases: a)~$\varphi_0={\pi}/{8}$, b)~$\varphi_0={\pi}/{4}$, c)~$\varphi_0={3 \pi}/{8}$.}
  \label{fig:G0vsV_difffi}
\end{figure}

Using the obtained formula for $g_{0}(\ep)$ and $\delta_{dc}g(\ep)$
we can calculate the conductance $G_{0}$ and its correction
$\delta G$ due to microwave radiation for different values of parameters
(damping $\ep _{N}$, phase difference $2\varphi _{0}$, etc.). The dependence of the conductance in the absence
of radiation $G_{0}$ versus the applied voltage $V$ is presented in
Fig.~\ref{fig:G0vsV_difffi}. We see that this dependence follows the energy
dependence of a SIN junction. In our case the $n$ wire with an induced subgap
plays a role of ''S'' with a damping $\ep_{N}$ in the ''superconductor''.
Since the value of the induced subgap, $\ep_{sg} = \ep_{S} |\cos \varphi_{0}|$,
depends on the phase difference $2 \varphi_{0}$, the position of the peak in
the dependence $G_{0}(V)$ is shifted downwards with increasing $\varphi_{0}$.
Note that in an asymmetrical system ($\ep_{S1}\neq \ep_{S2}$) the
lowest value of the subgap is not zero (cf. Eq.~\eqref{SPE8a}).\\

\begin{figure}
  \includegraphics[width = 8cm]{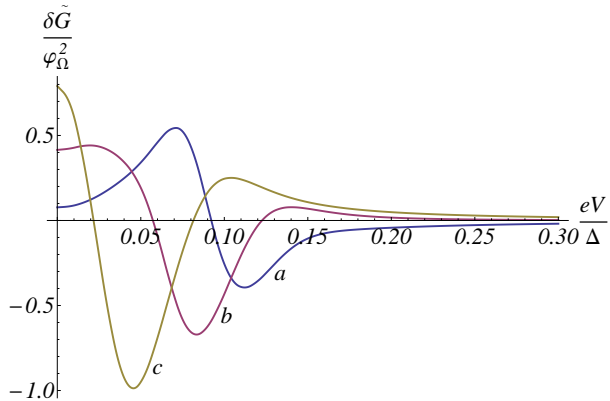}\\
  \caption{Normalized second-order correction of differential conductance $\delta \tilde{G}$ versus bias voltage $V$.
    Parameter values:
    $T/\Delta = 10^{-2}$, $\ep_S/\Delta=0.1$, $\ep_N/\Delta = 10^{-2}$, $\Omega/\Delta=5 \cdot 10^{-2}$.
    Different cases: a)~$\varphi_0={\pi}/{8}$, b)~$\varphi_0={\pi}/{4}$, c)~$\varphi_0={3 \pi}/{8}$.}
  \label{fig:dGvsV_difffi}
\end{figure}

In Fig.~\ref{fig:dGvsV_difffi} we show the bias voltage dependence of the conductance
correction due to ac radiation $\delta_{dc}G$ (coefficient in front of~$\varphi_{\Omega}^2$)
for different values of~$\varphi_0$.
The magnitude and the position of the arising peaks depend strongly on the values of the parameters,
e.g.~$\varphi_0$.\\

\begin{figure}
  \includegraphics[width = 8cm]{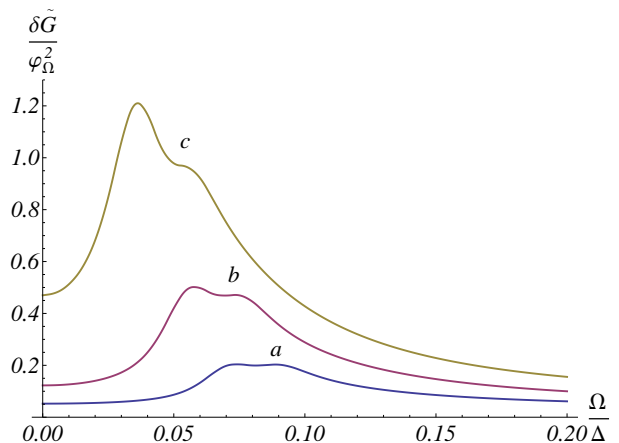}\\
  \caption{Normalized second-order correction of differential conductance $\delta \tilde{G}$ versus ac frequency $\Omega$.
    Parameter values: $T/\Delta=2\cdot 10^{-3}$, $\ep_S/\Delta=0.1$, $\ep_N/\Delta=10^{-2}$, $eV/\Delta=10^{-2}$.
    Different cases: a)~$\varphi_0={\pi}/{6}$, b)~$\varphi_0={\pi}/{4}$, c)~$\varphi_0={\pi}/{3}$.}
  \label{fig:dGvsOmega_difffi}
\end{figure}

\begin{figure}
  \includegraphics[width = 8cm]{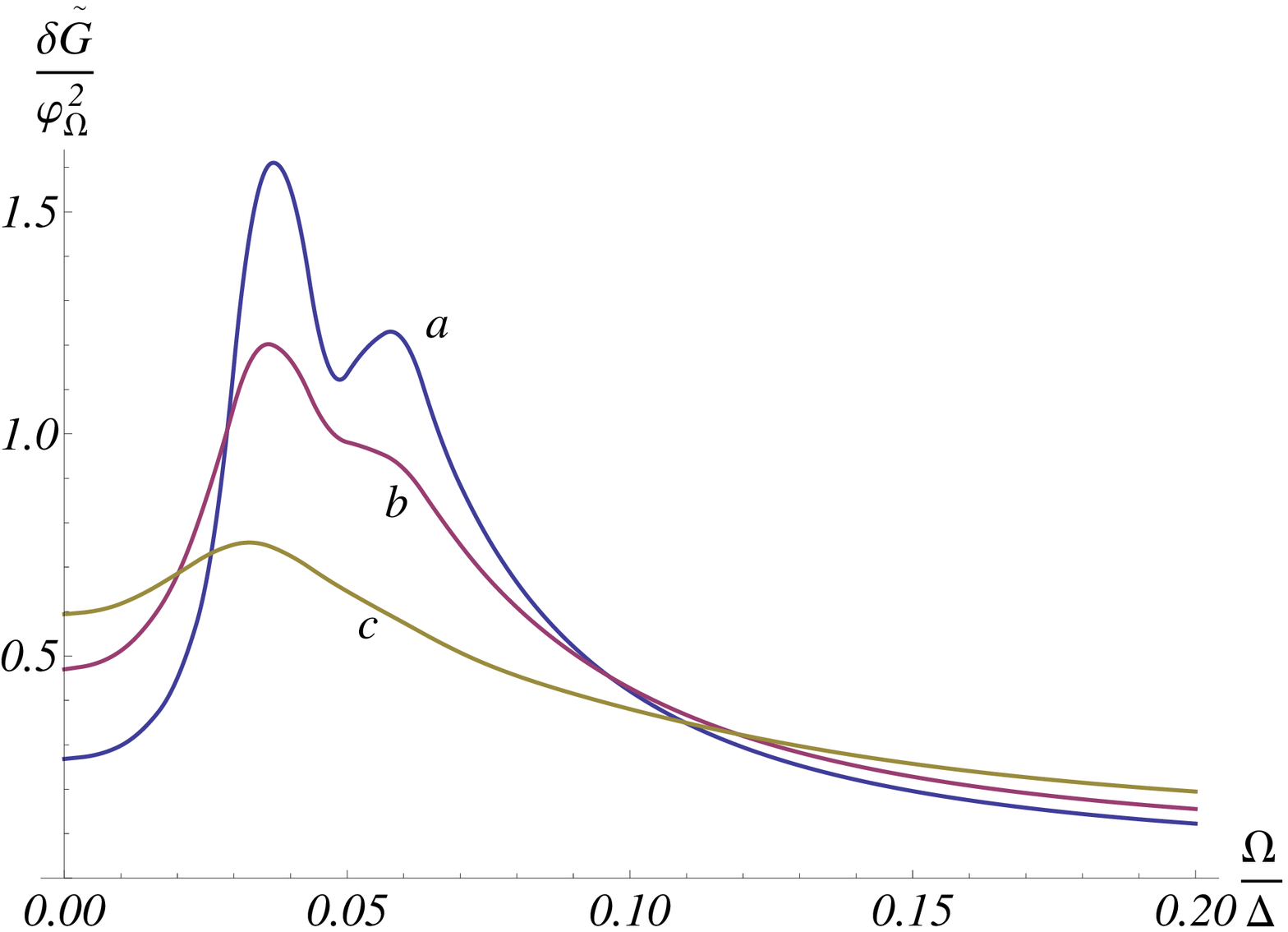}\\
  \caption{Normalized second-order correction of differential conductance $\delta \tilde{G}$ versus ac frequency $\Omega$.
    Parameter values: $T/\Delta=2\cdot 10^{-3}$, $\ep_S/\Delta=0.1$, $eV/\Delta=10^{-2}$, $\varphi_0=\pi/3$.
    Different cases: a)~$\ep_N/\Delta=5\cdot 10^{-3}$, b)~$\ep_N/\Delta=10^{-2}$, c)~$\ep_N/\Delta=2 \cdot 10^{-2}$.}
  \label{fig:dGvsOmega_diffDamp}
\end{figure}

\begin{figure}
  \includegraphics[width = 8cm]{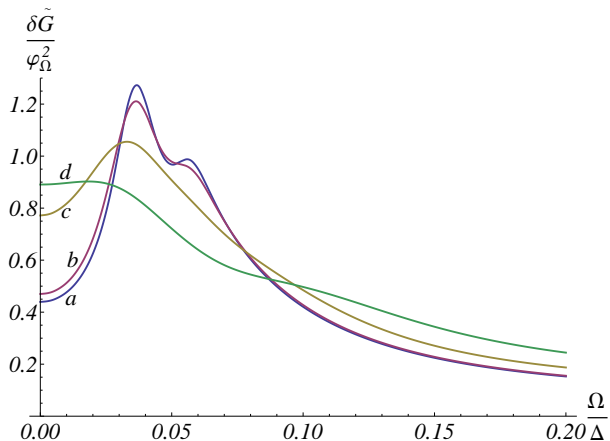}\\
  \caption{Normalized second-order correction of differential conductance $\delta \tilde{G}$ versus ac frequency $\Omega$,
    Parameter values: $\ep_S/\Delta=0.1$, $\ep_N/\Delta=10^{-2}$, $eV/\Delta=10^{-2}$, $\varphi_0=\pi/3$.
    Different cases: a)~$T=0$, b)~$T/\Delta=2\cdot 10^{-3}$, c)~$T/\Delta=6 \cdot 10^{-3}$, d)~$T/\Delta= 10^{-2}$.}
  \label{fig:dGvsOmega_diffT}
\end{figure}

By varying the stationary phase difference $\varphi_{0}$ or the damping $\ep_{N}$
one can change the frequency dependence of the correction $\delta_{dc}G$ considerably.
This is shown in Fig.~\ref{fig:dGvsOmega_difffi} and Fig.~\ref{fig:dGvsOmega_diffDamp} respectively.
One can see that if $\ep_{N} \ll \ep_{sg} (\varphi_{0})$, then the dependence
$\delta_{dc}G(\Omega)$ has a peak located at $\approx \ep_{sg}(\varphi_{0})$
and split into two subpeaks. The splitting becomes more and more distinct with increasing bias voltage V.
With decreasing $\varphi_{0}$ and increasing $\ep_{N}$, the form of this dependence changes
significantly. For example, the resonance curve becomes broader with increasing damping.
Increasing temperature leads to a similar effect as one can see in
Fig.~\ref{fig:dGvsOmega_diffT}.\\

\begin{figure}
  \includegraphics[width = 8cm]{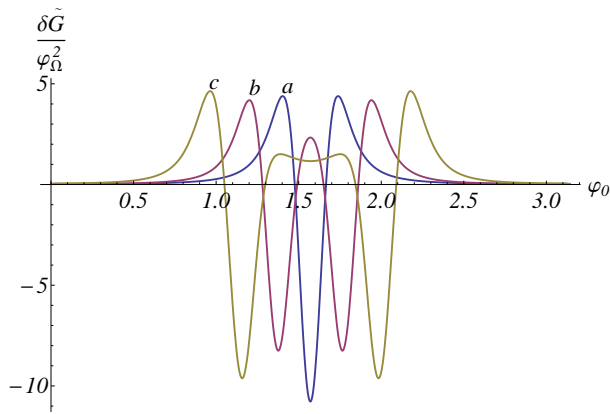}\\
  \caption{Normalized second-order correction of differential conductance $\delta \tilde{G}$ versus
    stationary phase difference $\varphi_0$.
    Parameter values: $T/\Delta=2\cdot 10^{-3}$, $\ep_S/\Delta=0.1$, $\ep_N/\Delta=10^{-2}$, $\Omega/\Delta=10^{-2}$.
    Different cases: a)~$V=0$, b)~$eV/\Delta= 2 \cdot 10^{-2}$, c)~$eV/\Delta=4 \cdot 10^{-2}$.}
  \label{fig:dGvsfi_diffV}
\end{figure}

In Fig. \ref{fig:dGvsfi_diffV} we plot the normalized conductance
correction~$\delta_{dc}\tilde{G}(\varphi_{0})$ as a function of $\varphi_0$ for different
values of the bias voltage $V$. At large $V$ this dependence is close to sinusoidal, but at
smaller voltages the form of the periodic function $\delta_{dc} G (\varphi_{0})$ becomes
more complicated.

\section{Conclusion}

We have calculated the change of the conductance in an S/N
structure of the cross geometry under the influence of microwave radiation.
The calculations have been carried out on the basis of quasiclassical
Green's functions in the diffusive limit. Two different limiting cases have
been considered: a) a weak proximity effect and low frequency $\Omega $ of
radiation; b) a strong proximity effect and small amplitude of radiation.\\

In the case a), the conductance change $\delta G$ consists of two parts. One
is related to a change of the $nN$ interface resistance due to a
modification of the DOS of the $n$ wire. At small applied voltages $V_{N},$
it is negative.  Another part is caused by a modification of the conductance
of the $n$ wire due to the PE. This part is positive and consists of two
competing contributions. One contribution, which is negative, stems from the
a modification of the DOS of the $n$ wire. Another contribution is positive
and caused by a term, which is similar to the Maki-Thompson term\cite
{Volkov&Pavl,Zaikin97}. The conductance change $\delta G$ oscillates and
decays with increasing amplitude of radiation.\\

In the case b) a short $n$ wire was considered so that the
resistance of the $n$ wire is negligible in comparison with the
resistance of the $nN$ interface. The correction $\delta G$ has
been found under assumption of a small amplitude of the radiation.
We found that at small applied voltages $V$, the dependence
$\delta G(\Omega )$ has a resonance
form. It has a maximum when the frequency $\Omega $ is of the order of
$\ep_{sg}=\ep_{S}|\cos \varphi_{0}|$ where $\ep_{sg}$ is a
subgap in the spectrum of the $n$ wire induced by the PE. With increasing $V$,
the peak in the dependence $\delta G(\Omega)$ splits into two peaks and
overall form of this dependence becomes complicated.\\

We assumed that the $nS$ interface resistance is larger
than the resistance of the $n$ wire, that is: $\rho L\ll R_{nS}$.
This inequality can be written in the form $\ep_{S}\ll
\ep_{Th} \equiv D/L^{2}$, that is, the subgap energy in the $n$
wire is much smaller than the Thouless energy  $\ep_{Th}$. In the
opposite limit, $\ep_{S}\gg \ep_{Th}$, a gap of the order of $\ep_{Th}$
is induced in the $n$ wire. This limit can be studied numerically.
However, one can expect that in this limit the resonance should take place at
$\omega_{res}\approx \ep_{Th}/\hbar$. Experiment performed in Ref.\cite{Petr10}
corresponds to this limit. The frequency corresponding to the Thouless
energy in experiment is equal to
$\ep_{Th}/h \approx \tfrac{1}{2\pi}(400/10^{-8})s^{-1}\approx 6 GHz$,
whereas the resonance frequency is $\nu_{res}\approx 10 GHz$.\\

Note that we considered a simplified model. For example, we did not
account for the change of the distribution function in the $n$
film (heating effects). One can give estimations when the ''heating'' can be
neglected. The change of an ''effective'' electron temperature $\delta T$
in the $n$ wire is approximately given by:
$\delta T\approx \tau_{e-ph}\sigma E^{2}/c_{e}$, where $\tau_{e-ph}$
is the electron-phonon inelastic scattering time,
$E\lesssim \delta V_{S} R_{L}/R_{b} L = \hbar \Omega (R_{L}/eL R_{b})
\varphi_{\Omega}$ is the ac electric field in the $n$ wire and
$c_{e}\approx T\cdot n/\ep_{F}$ is the heat capacity of electron
gas with concentration $n$. Taking into account that $(R_{b}\sigma)^{-1}\approx Z^{2}/l$,
we find that
$\delta T/T\lesssim (\ep_{0}/T)^{2}(\tau _{e-ph}/\tau ) Z^{4}\varphi_{\Omega }^{2}$,
where $Z^{2}$ is the dimensionless coefficient of electron penetration through the SN interface,
which is assumed to be small, $l=v\tau$ is the mean free path in the $n$ wire.
Therefore, the heating would be very small if the condition
$\varphi_{\Omega }\ll (\ep _{0}/T)\sqrt{(\tau /\tau _{e-ph})}Z^{-2}$ is fulfilled.\\

The obtained results are useful for understanding the response of
the considered and analogous SN systems to microwave radiation which can be
used, for example, in Q-bits.\\

We would like to thank SFB 491 for financial support. One of us (VTP) was
supported by the British EPSRC (Grant EP/F016891/1).

\section{Appendix}

The DOS in Eqs.~(\ref{WPEbc}-\ref{WPEj}) is given by the formula
$\nu (\ep,\varphi) = \Re [1 + \tfrac12 f(L_{x})^{2}]$ with $f(L_{x})$
defined in Eq.~\eqref{WPEfR}.\\

Making use of the weak proximity effect approximation the function we
rewrite $\langle M(\ep,\varphi)^{-1} \rangle $ in Eq.~\eqref{WPEj} as
\begin{equation}
    \langle M(\ep,\varphi)^{-1}\rangle =
        1 -  \tfrac12 \langle \Re \{ f^{2} + f f^{\ast} \} \rangle  \label{A1}
\end{equation}

    Using Eq.~\eqref{WPEfR} one can easily calculate
\begin{gather}
    \!\!\!\langle f^{2}\rangle =\frac{C^{2}+S^{2}}{2}\frac{\sinh 2\theta_{x} }{2\theta_{x}}
        +\frac{C^{2}-S^{2}}{2} + CS \frac{\sinh^2\theta_{x}}{\theta_{x}}  \label{A2}\\
    \langle ff^{\ast} \rangle = \frac{|C|^{2}+|S|^{2}}{2}\frac{\sinh 2\theta_{x}'}
        {2\theta_{x}'}+\frac{|C|^{2}-|S|^{2}}{2}\frac{\sin 2\theta_{x}''} {2\theta_{x}''}\nonumber\\
    \ \ +\Re\left\{C^{\ast}S \lb \frac{\sinh^2 \theta_{x}'}{\theta_{x}'}
        +i\frac{\sin^2 \theta_{x}''}{\theta_{x}''} \rb \right\} \label{A3}
\end{gather}
where $\theta_x'$ and $\theta_x''$ are the real and imaginary parts of $\theta_x$ respectively, i.e.
$\theta_x = \theta_x' + i \theta_x''$.\\

We use these expressions for calculating the function $\langle M(\ep,\varphi)^{-1}\rangle$
and conductance $G$.


\begin{thebibliography}{51}
\expandafter\ifx\csname natexlab\endcsname\relax\def\natexlab#1{#1}\fi
\expandafter\ifx\csname bibnamefont\endcsname\relax
  \def\bibnamefont#1{#1}\fi
\expandafter\ifx\csname bibfnamefont\endcsname\relax
  \def\bibfnamefont#1{#1}\fi
\expandafter\ifx\csname citenamefont\endcsname\relax
  \def\citenamefont#1{#1}\fi
\expandafter\ifx\csname url\endcsname\relax
  \def\url#1{\texttt{#1}}\fi
\expandafter\ifx\csname urlprefix\endcsname\relax\def\urlprefix{URL }\fi
\providecommand{\bibinfo}[2]{#2}
\providecommand{\eprint}[2][]{\url{#2}}

\bibitem[{\citenamefont{Petrashov et~al.}(1992)\citenamefont{Petrashov,
  Antonov, and Persson}}]{Petr92}
\bibinfo{author}{\bibfnamefont{V.~T.} \bibnamefont{Petrashov}},
  \bibinfo{author}{\bibfnamefont{V.~N.} \bibnamefont{Antonov}},
  \bibnamefont{and} \bibinfo{author}{\bibfnamefont{M.}~\bibnamefont{Persson}},
  \bibinfo{journal}{Physica Scripta} \textbf{\bibinfo{volume}{T42}},
  \bibinfo{pages}{136} (\bibinfo{year}{1992}), \bibinfo{note}{see also "Low
  Dimensional Properties of Solids" (Proceedings of the Nobel Jubilee
  Symposium, G\"oteborg, Sweden, December 4-7,1991) ed. M. Jonson and T.
  Claeson, World Scientific Publishing,1992}.

\bibitem[{\citenamefont{Petrashov et~al.}(1993)\citenamefont{Petrashov,
  Antonov, Delsing, and Claeson}}]{Petr93}
\bibinfo{author}{\bibfnamefont{V.~T.} \bibnamefont{Petrashov}},
  \bibinfo{author}{\bibfnamefont{V.~N.} \bibnamefont{Antonov}},
  \bibinfo{author}{\bibfnamefont{P.}~\bibnamefont{Delsing}}, \bibnamefont{and}
  \bibinfo{author}{\bibfnamefont{R.}~\bibnamefont{Claeson}},
  \bibinfo{journal}{Phys. Rev. Lett.} \textbf{\bibinfo{volume}{70}},
  \bibinfo{pages}{347} (\bibinfo{year}{1993}), \bibinfo{note}{ibid {\bf 74},
  5268 (1995)}.

\bibitem[{\citenamefont{Pothier et~al.}(1994)\citenamefont{Pothier, Gu\'eron,
  Esteve, and Devoret}}]{Pothier94}
\bibinfo{author}{\bibfnamefont{H.}~\bibnamefont{Pothier}},
  \bibinfo{author}{\bibfnamefont{S.}~\bibnamefont{Gu\'eron}},
  \bibinfo{author}{\bibfnamefont{D.}~\bibnamefont{Esteve}}, \bibnamefont{and}
  \bibinfo{author}{\bibfnamefont{M.~H.} \bibnamefont{Devoret}},
  \bibinfo{journal}{Phys. Rev. Lett.} \textbf{\bibinfo{volume}{73}},
  \bibinfo{pages}{2488} (\bibinfo{year}{1994}).

\bibitem[{\citenamefont{de~Vegvar et~al.}(1994)\citenamefont{de~Vegvar, Fulton,
  Mallison, and Miller}}]{Vegvar94}
\bibinfo{author}{\bibfnamefont{P.~G.~N.} \bibnamefont{de~Vegvar}},
  \bibinfo{author}{\bibfnamefont{T.~A.} \bibnamefont{Fulton}},
  \bibinfo{author}{\bibfnamefont{W.~H.} \bibnamefont{Mallison}},
  \bibnamefont{and} \bibinfo{author}{\bibfnamefont{R.~E.}
  \bibnamefont{Miller}}, \bibinfo{journal}{Phys. Rev. Lett.}
  \textbf{\bibinfo{volume}{73}}, \bibinfo{pages}{1416} (\bibinfo{year}{1994}).

\bibitem[{\citenamefont{Dimoulas et~al.}(1995)\citenamefont{Dimoulas, Heida,
  Wees, Klapwijk, Graaf, and Borghs}}]{Klapwijk95}
\bibinfo{author}{\bibfnamefont{A.}~\bibnamefont{Dimoulas}},
  \bibinfo{author}{\bibfnamefont{J.~P.} \bibnamefont{Heida}},
  \bibinfo{author}{\bibfnamefont{B.~J.~v.} \bibnamefont{Wees}},
  \bibinfo{author}{\bibfnamefont{T.~M.} \bibnamefont{Klapwijk}},
  \bibinfo{author}{\bibfnamefont{W.~v.~d.} \bibnamefont{Graaf}},
  \bibnamefont{and} \bibinfo{author}{\bibfnamefont{G.}~\bibnamefont{Borghs}},
  \bibinfo{journal}{Phys. Rev. Lett.} \textbf{\bibinfo{volume}{74}},
  \bibinfo{pages}{602} (\bibinfo{year}{1995}).

\bibitem[{\citenamefont{Eom et~al.}(1998)\citenamefont{Eom, Chien, and
  Chandrasekhar}}]{Chandra98}
\bibinfo{author}{\bibfnamefont{J.}~\bibnamefont{Eom}},
  \bibinfo{author}{\bibfnamefont{C.-J.} \bibnamefont{Chien}}, \bibnamefont{and}
  \bibinfo{author}{\bibfnamefont{V.}~\bibnamefont{Chandrasekhar}},
  \bibinfo{journal}{Phys. Rev. Lett.} \textbf{\bibinfo{volume}{81}},
  \bibinfo{pages}{437} (\bibinfo{year}{1998}).

\bibitem[{\citenamefont{Beenakker}(1995)}]{Been}
\bibinfo{author}{\bibfnamefont{C.~W.~J.} \bibnamefont{Beenakker}}, in
  \emph{\bibinfo{booktitle}{Mesoscopic Quantum Physics}}
  (\bibinfo{publisher}{North-Holland}, \bibinfo{address}{Amsterdam},
  \bibinfo{year}{1995}), \bibinfo{note}{ed. by E.~Akkermans, G.~Montambaux,
  J.-L.~Pichard and J.~Zinn-Justin}.

\bibitem[{\citenamefont{Lambert and Raimondi}(1998)}]{LambRaim}
\bibinfo{author}{\bibfnamefont{C.~J.} \bibnamefont{Lambert}} \bibnamefont{and}
  \bibinfo{author}{\bibfnamefont{R.}~\bibnamefont{Raimondi}},
  \bibinfo{journal}{J. Phys.: Condens. Matter} \textbf{\bibinfo{volume}{10}},
  \bibinfo{pages}{901} (\bibinfo{year}{1998}).

\bibitem[{\citenamefont{Nazarov}(1999)}]{NazRev}
\bibinfo{author}{\bibfnamefont{Y.~V.} \bibnamefont{Nazarov}},
  \bibinfo{journal}{Superlattices and Microstructures}
  \textbf{\bibinfo{volume}{25}}, \bibinfo{pages}{1221} (\bibinfo{year}{1999}).

\bibitem[{\citenamefont{Belzig et~al.}(1999)\citenamefont{Belzig, Wilhelm,
  Bruder, Sch\"on, and Zaikin}}]{Belzig}
\bibinfo{author}{\bibfnamefont{W.}~\bibnamefont{Belzig}},
  \bibinfo{author}{\bibfnamefont{F.~K.} \bibnamefont{Wilhelm}},
  \bibinfo{author}{\bibfnamefont{C.}~\bibnamefont{Bruder}},
  \bibinfo{author}{\bibfnamefont{G.}~\bibnamefont{Sch\"on}}, \bibnamefont{and}
  \bibinfo{author}{\bibfnamefont{A.~D.} \bibnamefont{Zaikin}},
  \bibinfo{journal}{Superlattices and Microstructures}
  \textbf{\bibinfo{volume}{25}}, \bibinfo{pages}{1251} (\bibinfo{year}{1999}).

\bibitem[{\citenamefont{Volkov et~al.}(1993)\citenamefont{Volkov, Zaitsev, and
  Klapwijk}}]{VZK}
\bibinfo{author}{\bibfnamefont{A.}~\bibnamefont{Volkov}},
  \bibinfo{author}{\bibfnamefont{A.}~\bibnamefont{Zaitsev}}, \bibnamefont{and}
  \bibinfo{author}{\bibfnamefont{T.}~\bibnamefont{Klapwijk}},
  \bibinfo{journal}{Physica C: Superconductivity}
  \textbf{\bibinfo{volume}{210}}, \bibinfo{pages}{21} (\bibinfo{year}{1993}).

\bibitem[{\citenamefont{Hekking and Nazarov}(1993)}]{Hekking93}
\bibinfo{author}{\bibfnamefont{F.~W.~J.} \bibnamefont{Hekking}}
  \bibnamefont{and} \bibinfo{author}{\bibfnamefont{Y.~V.}
  \bibnamefont{Nazarov}}, \bibinfo{journal}{Phys. Rev. Lett.}
  \textbf{\bibinfo{volume}{71}}, \bibinfo{pages}{1625} (\bibinfo{year}{1993}).

\bibitem[{\citenamefont{Zaitsev}(1994)}]{Zaitsev94}
\bibinfo{author}{\bibfnamefont{A.}~\bibnamefont{Zaitsev}},
  \bibinfo{journal}{Physica B: Condensed Matter}
  \textbf{\bibinfo{volume}{203}}, \bibinfo{pages}{274} (\bibinfo{year}{1994}).

\bibitem[{\citenamefont{Nazarov and Stoof}(1996)}]{Nazarov96}
\bibinfo{author}{\bibfnamefont{Y.~V.} \bibnamefont{Nazarov}} \bibnamefont{and}
  \bibinfo{author}{\bibfnamefont{T.~H.} \bibnamefont{Stoof}},
  \bibinfo{journal}{Phys. Rev. Lett.} \textbf{\bibinfo{volume}{76}},
  \bibinfo{pages}{823} (\bibinfo{year}{1996}).

\bibitem[{\citenamefont{Artemenko et~al.}(1979)\citenamefont{Artemenko, Volkov,
  and Zaitsev}}]{AVZ79}
\bibinfo{author}{\bibfnamefont{S.~N.} \bibnamefont{Artemenko}},
  \bibinfo{author}{\bibfnamefont{A.~F.} \bibnamefont{Volkov}},
  \bibnamefont{and} \bibinfo{author}{\bibfnamefont{A.~V.}
  \bibnamefont{Zaitsev}}, \bibinfo{journal}{Solid State Communications}
  \textbf{\bibinfo{volume}{30}}, \bibinfo{pages}{771} (\bibinfo{year}{1979}).

\bibitem[{\citenamefont{Volkov et~al.}(1996)\citenamefont{Volkov, Allsopp, and
  Lambert}}]{LambVolkov96}
\bibinfo{author}{\bibfnamefont{A.}~\bibnamefont{Volkov}},
  \bibinfo{author}{\bibfnamefont{N.}~\bibnamefont{Allsopp}}, \bibnamefont{and}
  \bibinfo{author}{\bibfnamefont{C.~J.} \bibnamefont{Lambert}},
  \bibinfo{journal}{Journal of Physics: Condensed Matter}
  \textbf{\bibinfo{volume}{8}}, \bibinfo{pages}{L45} (\bibinfo{year}{1996}).

\bibitem[{\citenamefont{Volkov}(1995)}]{Volkov95}
\bibinfo{author}{\bibfnamefont{A.~F.} \bibnamefont{Volkov}},
  \bibinfo{journal}{Phys. Rev. Lett.} \textbf{\bibinfo{volume}{74}},
  \bibinfo{pages}{4730} (\bibinfo{year}{1995}).

\bibitem[{\citenamefont{Wilhelm et~al.}(1998)\citenamefont{Wilhelm, Sch\"on,
  and Zaikin}}]{Schon98}
\bibinfo{author}{\bibfnamefont{F.~K.} \bibnamefont{Wilhelm}},
  \bibinfo{author}{\bibfnamefont{G.}~\bibnamefont{Sch\"on}}, \bibnamefont{and}
  \bibinfo{author}{\bibfnamefont{A.~D.} \bibnamefont{Zaikin}},
  \bibinfo{journal}{Phys. Rev. Lett.} \textbf{\bibinfo{volume}{81}},
  \bibinfo{pages}{1682} (\bibinfo{year}{1998}).

\bibitem[{\citenamefont{Yip}(1998)}]{Yip98}
\bibinfo{author}{\bibfnamefont{S.-K.} \bibnamefont{Yip}},
  \bibinfo{journal}{Phys. Rev. B} \textbf{\bibinfo{volume}{58}},
  \bibinfo{pages}{5803} (\bibinfo{year}{1998}).

\bibitem[{\citenamefont{Gubankov and Margolin}(1979)}]{Marg79}
\bibinfo{author}{\bibfnamefont{V.~N.} \bibnamefont{Gubankov}} \bibnamefont{and}
  \bibinfo{author}{\bibfnamefont{N.~M.} \bibnamefont{Margolin}},
  \bibinfo{journal}{JETP Lett.} \textbf{\bibinfo{volume}{29}},
  \bibinfo{pages}{673} (\bibinfo{year}{1979}).

\bibitem[{\citenamefont{Charlat et~al.}(1996)\citenamefont{Charlat, Courtois,
  Gandit, Mailly, Volkov, and Pannetier}}]{Pannetier96}
\bibinfo{author}{\bibfnamefont{P.}~\bibnamefont{Charlat}},
  \bibinfo{author}{\bibfnamefont{H.}~\bibnamefont{Courtois}},
  \bibinfo{author}{\bibfnamefont{P.}~\bibnamefont{Gandit}},
  \bibinfo{author}{\bibfnamefont{D.}~\bibnamefont{Mailly}},
  \bibinfo{author}{\bibfnamefont{A.~F.} \bibnamefont{Volkov}},
  \bibnamefont{and}
  \bibinfo{author}{\bibfnamefont{B.}~\bibnamefont{Pannetier}},
  \bibinfo{journal}{Phys. Rev. Lett.} \textbf{\bibinfo{volume}{77}},
  \bibinfo{pages}{4950} (\bibinfo{year}{1996}).

\bibitem[{\citenamefont{Petrashov et~al.}(1996)\citenamefont{Petrashov,
  Shaikhaidarov, and Sosnin}}]{Petr96}
\bibinfo{author}{\bibfnamefont{V.}~\bibnamefont{Petrashov}},
  \bibinfo{author}{\bibfnamefont{R.}~\bibnamefont{Shaikhaidarov}},
  \bibnamefont{and} \bibinfo{author}{\bibfnamefont{I.}~\bibnamefont{Sosnin}},
  \bibinfo{journal}{JETP Letters} \textbf{\bibinfo{volume}{64}},
  \bibinfo{pages}{839} (\bibinfo{year}{1996}).

\bibitem[{\citenamefont{Baselmans et~al.}(1999)\citenamefont{Baselmans,
  Morpurgo, van Wees, and Klapwijk}}]{KlapWees}
\bibinfo{author}{\bibfnamefont{J.~J.~A.} \bibnamefont{Baselmans}},
  \bibinfo{author}{\bibfnamefont{A.~F.} \bibnamefont{Morpurgo}},
  \bibinfo{author}{\bibfnamefont{B.~J.} \bibnamefont{van Wees}},
  \bibnamefont{and} \bibinfo{author}{\bibfnamefont{T.~M.}
  \bibnamefont{Klapwijk}}, \bibinfo{journal}{Nature}
  \textbf{\bibinfo{volume}{397}}, \bibinfo{pages}{43} (\bibinfo{year}{1999}).

\bibitem[{\citenamefont{Shaikhaidarov et~al.}(2000)\citenamefont{Shaikhaidarov,
  Volkov, Takayanagi, Petrashov, and Delsing}}]{Petr00}
\bibinfo{author}{\bibfnamefont{R.}~\bibnamefont{Shaikhaidarov}},
  \bibinfo{author}{\bibfnamefont{A.~F.} \bibnamefont{Volkov}},
  \bibinfo{author}{\bibfnamefont{H.}~\bibnamefont{Takayanagi}},
  \bibinfo{author}{\bibfnamefont{V.~T.} \bibnamefont{Petrashov}},
  \bibnamefont{and} \bibinfo{author}{\bibfnamefont{P.}~\bibnamefont{Delsing}},
  \bibinfo{journal}{Phys. Rev. B} \textbf{\bibinfo{volume}{62}},
  \bibinfo{pages}{R14649} (\bibinfo{year}{2000}).

\bibitem[{\citenamefont{Golubov et~al.}(2004)\citenamefont{Golubov, Kupriyanov,
  and Il\char39{}ichev}}]{GolubovRMP}
\bibinfo{author}{\bibfnamefont{A.~A.} \bibnamefont{Golubov}},
  \bibinfo{author}{\bibfnamefont{M.~Y.} \bibnamefont{Kupriyanov}},
  \bibnamefont{and}
  \bibinfo{author}{\bibfnamefont{E.}~\bibnamefont{Il\char39{}ichev}},
  \bibinfo{journal}{Rev. Mod. Phys.} \textbf{\bibinfo{volume}{76}},
  \bibinfo{pages}{411} (\bibinfo{year}{2004}).

\bibitem[{\citenamefont{Buzdin}(2005)}]{BuzdinRMP}
\bibinfo{author}{\bibfnamefont{A.~I.} \bibnamefont{Buzdin}},
  \bibinfo{journal}{Rev. Mod. Phys.} \textbf{\bibinfo{volume}{77}},
  \bibinfo{pages}{935} (\bibinfo{year}{2005}).

\bibitem[{\citenamefont{Bergeret et~al.}(2005)\citenamefont{Bergeret, Volkov,
  and Efetov}}]{BVERMP}
\bibinfo{author}{\bibfnamefont{F.~S.} \bibnamefont{Bergeret}},
  \bibinfo{author}{\bibfnamefont{A.~F.} \bibnamefont{Volkov}},
  \bibnamefont{and} \bibinfo{author}{\bibfnamefont{K.~B.}
  \bibnamefont{Efetov}}, \bibinfo{journal}{Rev. Mod. Phys.}
  \textbf{\bibinfo{volume}{77}}, \bibinfo{pages}{1321} (\bibinfo{year}{2005}).

\bibitem[{\citenamefont{Tsuei and Kirtley}(2000)}]{Kirtley}
\bibinfo{author}{\bibfnamefont{C.~C.} \bibnamefont{Tsuei}} \bibnamefont{and}
  \bibinfo{author}{\bibfnamefont{J.~R.} \bibnamefont{Kirtley}},
  \bibinfo{journal}{Rev. Mod. Phys.} \textbf{\bibinfo{volume}{72}},
  \bibinfo{pages}{969} (\bibinfo{year}{2000}).

\bibitem[{\citenamefont{Van~Harlingen}(1995)}]{vHarlingen}
\bibinfo{author}{\bibfnamefont{D.~J.} \bibnamefont{Van~Harlingen}},
  \bibinfo{journal}{Rev. Mod. Phys.} \textbf{\bibinfo{volume}{67}},
  \bibinfo{pages}{515} (\bibinfo{year}{1995}).

\bibitem[{\citenamefont{Bezryadin et~al.}(2000)\citenamefont{Bezryadin, Lau,
  and Tinkham}}]{Tinkham00}
\bibinfo{author}{\bibfnamefont{A.}~\bibnamefont{Bezryadin}},
  \bibinfo{author}{\bibfnamefont{C.~N.} \bibnamefont{Lau}}, \bibnamefont{and}
  \bibinfo{author}{\bibfnamefont{M.}~\bibnamefont{Tinkham}},
  \bibinfo{journal}{Nature} \textbf{\bibinfo{volume}{404}},
  \bibinfo{pages}{971} (\bibinfo{year}{2000}).

\bibitem[{\citenamefont{Zgirski et~al.}(2005)\citenamefont{Zgirski, Riikonen,
  Touboltsev, and Arutyunov}}]{Arut05}
\bibinfo{author}{\bibfnamefont{M.}~\bibnamefont{Zgirski}},
  \bibinfo{author}{\bibfnamefont{K.-P.} \bibnamefont{Riikonen}},
  \bibinfo{author}{\bibfnamefont{V.}~\bibnamefont{Touboltsev}},
  \bibnamefont{and}
  \bibinfo{author}{\bibfnamefont{K.}~\bibnamefont{Arutyunov}},
  \bibinfo{journal}{Nano Letters} \textbf{\bibinfo{volume}{5}},
  \bibinfo{pages}{1029} (\bibinfo{year}{2005}).

\bibitem[{\citenamefont{Tian et~al.}(2005)\citenamefont{Tian, Kumar, Xu, Wang,
  Kurtz, and Chan}}]{Chan05}
\bibinfo{author}{\bibfnamefont{M.}~\bibnamefont{Tian}},
  \bibinfo{author}{\bibfnamefont{N.}~\bibnamefont{Kumar}},
  \bibinfo{author}{\bibfnamefont{S.}~\bibnamefont{Xu}},
  \bibinfo{author}{\bibfnamefont{J.}~\bibnamefont{Wang}},
  \bibinfo{author}{\bibfnamefont{J.~S.} \bibnamefont{Kurtz}}, \bibnamefont{and}
  \bibinfo{author}{\bibfnamefont{M.~H.~W.} \bibnamefont{Chan}},
  \bibinfo{journal}{Phys. Rev. Lett.} \textbf{\bibinfo{volume}{95}},
  \bibinfo{pages}{076802} (\bibinfo{year}{2005}).

\bibitem[{\citenamefont{Arutyunov et~al.}(2008)\citenamefont{Arutyunov,
  Golubev, and Zaikin}}]{Arut08}
\bibinfo{author}{\bibfnamefont{K.}~\bibnamefont{Arutyunov}},
  \bibinfo{author}{\bibfnamefont{D.}~\bibnamefont{Golubev}}, \bibnamefont{and}
  \bibinfo{author}{\bibfnamefont{A.}~\bibnamefont{Zaikin}},
  \bibinfo{journal}{Physics Reports} \textbf{\bibinfo{volume}{464}},
  \bibinfo{pages}{1} (\bibinfo{year}{2008}).

\bibitem[{\citenamefont{Petrashov et~al.}(2005)\citenamefont{Petrashov, Chua,
  Marshall, Shaikhaidarov, and Nicholls}}]{Petr05}
\bibinfo{author}{\bibfnamefont{V.~T.} \bibnamefont{Petrashov}},
  \bibinfo{author}{\bibfnamefont{K.~G.} \bibnamefont{Chua}},
  \bibinfo{author}{\bibfnamefont{K.~M.} \bibnamefont{Marshall}},
  \bibinfo{author}{\bibfnamefont{R.~S.} \bibnamefont{Shaikhaidarov}},
  \bibnamefont{and} \bibinfo{author}{\bibfnamefont{J.~T.}
  \bibnamefont{Nicholls}}, \bibinfo{journal}{Phys. Rev. Lett.}
  \textbf{\bibinfo{volume}{95}}, \bibinfo{pages}{147001}
  (\bibinfo{year}{2005}).

\bibitem[{\citenamefont{Chiodi et~al.}(2009)\citenamefont{Chiodi, Aprili, and
  Reulet}}]{Aprili09}
\bibinfo{author}{\bibfnamefont{F.}~\bibnamefont{Chiodi}},
  \bibinfo{author}{\bibfnamefont{M.}~\bibnamefont{Aprili}}, \bibnamefont{and}
  \bibinfo{author}{\bibfnamefont{B.}~\bibnamefont{Reulet}}
  (\bibinfo{year}{2009}), \bibinfo{note}{arXiv:0908.1070}.

\bibitem[{\citenamefont{Checkley et~al.}(2010)\citenamefont{Checkley, Iagallo,
  Shaikhaidarov, Nicholls, and Petrashov}}]{Petr10}
\bibinfo{author}{\bibfnamefont{C.}~\bibnamefont{Checkley}},
  \bibinfo{author}{\bibfnamefont{A.}~\bibnamefont{Iagallo}},
  \bibinfo{author}{\bibfnamefont{R.}~\bibnamefont{Shaikhaidarov}},
  \bibinfo{author}{\bibfnamefont{J.~T.} \bibnamefont{Nicholls}},
  \bibnamefont{and} \bibinfo{author}{\bibfnamefont{V.~T.}
  \bibnamefont{Petrashov}} (\bibinfo{year}{2010}),
  \bibinfo{note}{arXiv:1003.2176}.

\bibitem[{\citenamefont{Golubev and Zaikin}(2009)}]{Zaikin09}
\bibinfo{author}{\bibfnamefont{D.}~\bibnamefont{Golubev}} \bibnamefont{and}
  \bibinfo{author}{\bibfnamefont{A.}~\bibnamefont{Zaikin}},
  \bibinfo{journal}{Europhys. Lett.} \textbf{\bibinfo{volume}{86}},
  \bibinfo{pages}{37009} (\bibinfo{year}{2009}).

\bibitem[{\citenamefont{Virtanen et~al.}(2010)\citenamefont{Virtanen,
  Heikkil\"{a}, Bergeret, and Cuevas}}]{Bergeret10}
\bibinfo{author}{\bibfnamefont{P.}~\bibnamefont{Virtanen}},
  \bibinfo{author}{\bibfnamefont{T.~T.} \bibnamefont{Heikkil\"{a}}},
  \bibinfo{author}{\bibfnamefont{F.~S.} \bibnamefont{Bergeret}},
  \bibnamefont{and} \bibinfo{author}{\bibfnamefont{J.~C.} \bibnamefont{Cuevas}}
  (\bibinfo{year}{2010}), \bibinfo{note}{arXiv:1001.5149}.

\bibitem[{\citenamefont{Giazotto et~al.}(2006)\citenamefont{Giazotto, Heikkila,
  Luukanen, Savin, and Pekola}}]{Pekola}
\bibinfo{author}{\bibfnamefont{F.}~\bibnamefont{Giazotto}},
  \bibinfo{author}{\bibfnamefont{T.~T.} \bibnamefont{Heikkila}},
  \bibinfo{author}{\bibfnamefont{A.}~\bibnamefont{Luukanen}},
  \bibinfo{author}{\bibfnamefont{A.~M.} \bibnamefont{Savin}}, \bibnamefont{and}
  \bibinfo{author}{\bibfnamefont{J.~P.} \bibnamefont{Pekola}},
  \bibinfo{journal}{Rev. Mod. Phys.} \textbf{\bibinfo{volume}{78}},
  \bibinfo{pages}{217} (\bibinfo{year}{2006}).

\bibitem[{\citenamefont{Usadel}(1970)}]{Usadel}
\bibinfo{author}{\bibfnamefont{K.~D.} \bibnamefont{Usadel}},
  \bibinfo{journal}{Phys. Rev. Lett.} \textbf{\bibinfo{volume}{25}},
  \bibinfo{pages}{507} (\bibinfo{year}{1970}).

\bibitem[{\citenamefont{Larkin and Ovchinnikov}(1969)}]{LO}
\bibinfo{author}{\bibfnamefont{A.~I.} \bibnamefont{Larkin}} \bibnamefont{and}
  \bibinfo{author}{\bibfnamefont{Y.~N.} \bibnamefont{Ovchinnikov}},
  \bibinfo{journal}{Sov. Phys. JETP} \textbf{\bibinfo{volume}{28}},
  \bibinfo{pages}{1200} (\bibinfo{year}{1969}), \bibinfo{note}{{\rm [Zh. Eksp.
  Teor. Fiz. {\bf 55}, 2262 (1968)]}}.

\bibitem[{\citenamefont{Rammer and Smith}(1986)}]{RS}
\bibinfo{author}{\bibfnamefont{J.}~\bibnamefont{Rammer}} \bibnamefont{and}
  \bibinfo{author}{\bibfnamefont{H.}~\bibnamefont{Smith}},
  \bibinfo{journal}{Rev. Mod. Phys.} \textbf{\bibinfo{volume}{58}},
  \bibinfo{pages}{323} (\bibinfo{year}{1986}).

\bibitem[{\citenamefont{Kopnin}(2001)}]{Kopnin}
\bibinfo{author}{\bibfnamefont{N.~B.} \bibnamefont{Kopnin}},
  \emph{\bibinfo{title}{{Theory of Nonequilibrium Superconductivity}}}
  (\bibinfo{publisher}{Clarendon Press}, \bibinfo{address}{Oxford},
  \bibinfo{year}{2001}).

\bibitem[{\citenamefont{Kogan et~al.}(2002)\citenamefont{Kogan, Pavlovskii, and
  Volkov}}]{VKogan}
\bibinfo{author}{\bibfnamefont{V.~R.} \bibnamefont{Kogan}},
  \bibinfo{author}{\bibfnamefont{V.~V.} \bibnamefont{Pavlovskii}},
  \bibnamefont{and} \bibinfo{author}{\bibfnamefont{A.~F.}
  \bibnamefont{Volkov}}, \bibinfo{journal}{Europhys. Lett.}
  \textbf{\bibinfo{volume}{59}}, \bibinfo{pages}{875} (\bibinfo{year}{2002}).

\bibitem[{\citenamefont{Volkov and Pavlovskii}(1996)}]{Volkov&Pavl}
\bibinfo{author}{\bibfnamefont{A.~F.} \bibnamefont{Volkov}} \bibnamefont{and}
  \bibinfo{author}{\bibfnamefont{V.~V.} \bibnamefont{Pavlovskii}}, in
  \emph{\bibinfo{booktitle}{Proceedings of the 21st Rencontres de Moriond}}
  (\bibinfo{address}{Les Arcs, France}, \bibinfo{year}{1996}), p.
  \bibinfo{pages}{267}.

\bibitem[{\citenamefont{Golubov et~al.}(1997)\citenamefont{Golubov, Wilhelm,
  and Zaikin}}]{Zaikin97}
\bibinfo{author}{\bibfnamefont{A.~A.} \bibnamefont{Golubov}},
  \bibinfo{author}{\bibfnamefont{F.~K.} \bibnamefont{Wilhelm}},
  \bibnamefont{and} \bibinfo{author}{\bibfnamefont{A.~D.}
  \bibnamefont{Zaikin}}, \bibinfo{journal}{Phys. Rev. B}
  \textbf{\bibinfo{volume}{55}}, \bibinfo{pages}{1123} (\bibinfo{year}{1997}).

\bibitem[{\citenamefont{Artemenko and Volkov}(1979)}]{ArtVolkov}
\bibinfo{author}{\bibfnamefont{S.~N.} \bibnamefont{Artemenko}}
  \bibnamefont{and} \bibinfo{author}{\bibfnamefont{A.~F.}
  \bibnamefont{Volkov}}, \bibinfo{journal}{Soviet Physics Uspekhi}
  \textbf{\bibinfo{volume}{22}}, \bibinfo{pages}{295} (\bibinfo{year}{1979}).

\bibitem[{\citenamefont{Larkin and Ovchinnikov}(1975)}]{LO75}
\bibinfo{author}{\bibfnamefont{A.~I.} \bibnamefont{Larkin}} \bibnamefont{and}
  \bibinfo{author}{\bibfnamefont{Y.~N.} \bibnamefont{Ovchinnikov}},
  \bibinfo{journal}{Sov. Phys. JETP} \textbf{\bibinfo{volume}{41}},
  \bibinfo{pages}{960} (\bibinfo{year}{1975}), \bibinfo{note}{{\rm [Zh. Eksp.
  Teor. Fiz. {\bf 68}, 1915 (1975)]}}.

\bibitem[{\citenamefont{Kupriyanov and Lukichev}(1988)}]{KL}
\bibinfo{author}{\bibfnamefont{M.~Y.} \bibnamefont{Kupriyanov}}
  \bibnamefont{and} \bibinfo{author}{\bibfnamefont{V.~F.}
  \bibnamefont{Lukichev}}, \bibinfo{journal}{Sov. Phys. JETP}
  \textbf{\bibinfo{volume}{67}}, \bibinfo{pages}{1163} (\bibinfo{year}{1988}),
  \bibinfo{note}{{\rm [Zh. Eksp. Teor. Fiz. {\bf 94}, 139 (1988)]}}.

\bibitem[{\citenamefont{McMillan}(1968)}]{McMillan}
\bibinfo{author}{\bibfnamefont{W.~L.} \bibnamefont{McMillan}},
  \bibinfo{journal}{Phys. Rev.} \textbf{\bibinfo{volume}{175}},
  \bibinfo{pages}{537} (\bibinfo{year}{1968}).

\bibitem[{\citenamefont{Zaitsev et~al.}(1999)\citenamefont{Zaitsev, Volkov,
  Bailey, and Lambert}}]{Zaitsev99}
\bibinfo{author}{\bibfnamefont{A.~V.} \bibnamefont{Zaitsev}},
  \bibinfo{author}{\bibfnamefont{A.~F.} \bibnamefont{Volkov}},
  \bibinfo{author}{\bibfnamefont{S.~W.~D.} \bibnamefont{Bailey}},
  \bibnamefont{and} \bibinfo{author}{\bibfnamefont{C.~J.}
  \bibnamefont{Lambert}}, \bibinfo{journal}{Phys. Rev. B}
  \textbf{\bibinfo{volume}{60}}, \bibinfo{pages}{3559} (\bibinfo{year}{1999}).

\end{thebibliography}
\end{document}